%% file: main.tex
\documentclass[preprint,journal]{vgtc}            

\newcommand{\eg}{{e.g.,}}
\usepackage[svgnames]{xcolor}
\usepackage[inkscapelatex=false]{svg}
\usepackage{amsmath}
\usepackage{soulutf8}

\newcommand{\revision}[1]{\textcolor{black}{#1}}

\definecolor{sampling}{HTML}{0E7946}
\definecolor{summary}{HTML}{870AB3}
\definecolor{functional}{HTML}{E04418}

\onlineid{1255}

\vgtccategory{Research}

\vgtcpapertype{Theoretical \& Empirical}

\title{Designing for Disclosure in Data Visualizations}

\author{%
  Krisha Mehta, Gordon Kindlmann and 
  Alex Kale
}

\authorfooter{
  \item
  	Krisha Mehta is with the University of Chicago.
  	E-mail: krisha@uchicago.edu.
  \item
  	Gordon Kindlmann is with the University of Chicago.
  	E-mail: glk@cs.uchicago.edu.

  \item Alex Kale is with the University of Chicago.
  	E-mail: kalea@uchicago.edu.
}

\abstract{%
 Visualizing data often entails data transformations that can reveal and hide information, operations we dub disclosure tactics. Whether designers hide information intentionally or as an implicit consequence of other design choices, tools and frameworks for visualization offer little explicit guidance on disclosure. To systematically characterize how visualizations can limit access to an underlying dataset, we contribute a content analysis of 425 examples of visualization techniques sampled from academic papers in the visualization literature, resulting in a taxonomy of disclosure tactics. 
 Our taxonomy organizes disclosure tactics based on how they change the data representation underlying a chart, providing a systematic way to reason about design trade-offs in terms of what information is revealed, distorted, or hidden.
 We demonstrate the benefits of using our taxonomy by showing how it can guide reasoning in design scenarios where disclosure is a first-order consideration. 
 Adopting disclosure as a framework for visualization research offers new perspective on authoring tools, literacy, uncertainty communication, personalization, and ethical design.
}


\teaser{
  \centering
  \includesvg[width=\linewidth]{images/comic}
  \setlength{\abovecaptionskip}{-3mm}
  \setlength{\belowcaptionskip}{-2.15mm}
  \caption{%
  An illustration of a privacy dilemma that requires using visualization as a disclosure mechanism.
  }
  \label{fig:teaser}
}

\graphicspath{{figs/}{figures/}{pictures/}{images/}{./}} 

\usepackage{tabu}                      
\usepackage{booktabs}                  
\usepackage{lipsum}                    
\usepackage{mwe}                       

\usepackage{mathptmx}                  

\begin{document}



\maketitle
\input{1_intro}
\input{2_relatedwork}
\input{3_methods}
\input{4_results}
\input{5_implications}
\input{6_discussion}

\bibliographystyle{abbrv-doi-hyperref}

\bibliography{main}

\end{document}

%% file: 1_intro.tex
\section{Introduction}
\label{intro}

Every time designers render a visualization, they must make choices about information disclosure\revision{:} 
what information is hidden and what is revealed in the data visualization. These choices create what we call a \textit{disclosure gap}, i.e., the difference between the information available in the original dataset and the information present in its corresponding visualization. 
The visualization community is already familiar with the issue of disclosure through concepts like Simpson's paradox \cite{simpson1951interpretation} and the ``modifiable areal unit problem'' \cite{fotheringham1991modifiable} that highlight how patterns in a visualization change depending on how the data are grouped.  Similarly, Shneiderman’s mantra, ``Overview first, zoom and filter, then details-on-demand,'' \cite{shneiderman2003eyes} calls to attention the level of detail revealed through a chart. 
However, we still lack the kind of systematic framework that would enable us to author visualizations with disclosure as a first-order consideration. 
\revision{As argued by recent work}~\cite{Wu2024-transforms}\revision{, the grammar of graphics determines exactly what information will be revealed by a visualization through both encoding and data pre-processing operations, thus offering a poor abstraction for reasoning about disclosure, whether a designer wishes to make intentional disclosure gaps or avoid inadvertent ones. For this reason, we find it necessary to distinguish between what we dub \textit{disclosure tactics}, data operations that lead to disclosure gaps, and \textit{emphasis tactics}, rendering choices such as mark types, scales, and layouts that do not influence disclosure. In this work, we explore how this distinction---and in particular renewed attention to disclosure---can benefit visualization research and practice.}

Consider the scenario presented in Figure \ref{fig:teaser}.
As a bio-statistician at a university medical center, Fred must answer queries from industry partners while protecting patient privacy.
In order to avoid disclosing individual patient data, Fred relies on visualizations with intentional disclosure gaps that selectively reveal and hide specific task-relevant pieces of information we call \textit{signals}.
Drug discovery, for example, may benefit from finding clusters in a relationship between patient genetics and treatment outcomes. Fred experiments with ways to reveal clusters while maintaining individual privacy:
(1) A heatmap could be used to alias together individuals similar to k-anonymity~\cite{sweeney2002-kanon}, but an out-of-the-box approach either gives so many bins as to reveal individuals or so few bins as to hide clusters.
(2) Differential privacy~\cite{panavas2023investigatingDiffPrivate, nanayakkara2022-dpTradeoffs} could be used to add noise to individual records, however, doing so in a way that preserves the number, location, and extent of clusters requires effort to fine-tune a bespoke algorithm.
(3) In contrast, a contour plot, created by deriving containment intervals from kernel density estimation, can produce the desired result with minimal effort on Fred's part.
In practice, Fred's ad hoc design process depends on guessing and checking possible solutions.
Section \ref{section5} revisits Fred's scenario, and similar design challenges, to demonstrate how a systematic framework for disclosure can empower designers to reason through the \textit{flexibility} of different visualizations to reveal and hide specific signals.

To solve Fred’s dilemma, and understand the issue of disclosure in visualization, we must 
\revision{develop theoretical frameworks that enable tool builders and designers to reason more proactively about disclosure tactics and their consequences for users of visualizations. The closest existing framework is \textit{algebraic visualization design} (AVD), which offers an account of different ways that visualizations can be vulnerable to miscommunicate due to non-correspondence between data and visualization. However, AVD does not describe the kinds of operations that lead to these vulnerabilities and thus offers limited support for explaining or planning around them. Our work takes an initial step toward such theory building by investigating (1) the variety of} disclosure tactics that can be used by designers to induce disclosure gaps, either deliberately or inadvertently, \revision{and (2) how developing a descriptive vocabulary about disclosure might} enable reasoning about design trade-offs around what information is revealed, distorted, or hidden.
\looseness=-1 

We contribute a taxonomy of disclosure tactics commonly used in visualization techniques, derived from a qualitative analysis of a diverse corpus of 425 examples sampled from the visualization research literature.
To outline the variety of operations that induce disclosure gaps between data and its visualization, we scope our qualitative analysis to non-interactive, 2D visualizations of tabular data.
\revision{We present 
composition patterns describing
how multiple disclosure tactics
can be
used together to render a visualization.}
We then characterize the implications of disclosure for visualization interpretation in terms of 
\revision{AVD vulnerabilities}~\cite{kindlmann2014-AVD}\revision{, following the way prior visualization research has applied this framework}~\cite{McNutt2020-Mirages, correll2023-terubozu, long2024-cutYaxis, correll2019-LooksGoodToMe}.
We demonstrate the practical value of our taxonomy via three design scenarios (e.g., Fred's privacy problem), showing how knowledge of disclosure tactics enables reasoning through trade-offs around disclosing task-specific information (i.e., \textit{signals}).
Our investigation lays the groundwork for new ways of thinking about disclosure more explicitly in visualization authoring and interpretation, with implications for topics such as visualization literacy, uncertainty visualization, personalization, and ethical design.

\begin{figure*}
  \centering
  \includesvg[width=\textwidth]{images/methods_flowtable.svg}
  \setlength{\abovecaptionskip}{-3.5mm}
  \setlength{\belowcaptionskip}{-7.45mm}
  \caption{A step-by-step description of how we built our corpus for analysis in two phases.}
\label{fig:flowchart}
\end{figure*}

%% file: 2_relatedwork.tex
\section{Background: Disclosure in Visualization}

Prior work in visualization 
discusses the distinction between what information is contained in a chart 
vs. how it
is presented.
In his seminal book, Bertin~\cite{bertin1983-semiology} highlighted the need to strictly separate the content (i.e., the information to be transmitted) from the container (i.e., the properties of the graphical system). 
Similarly, Mackinlay's~\cite{Mackinlay1986-AutoDesign} expressiveness and effectiveness criteria 
separate the need to encode the facts of a dataset with fidelity from the need to encode them in ways that will enable perceptually accurate decoding. 
Although the expressiveness criterion does not strictly isolate disclosure choices, it sets a broad imperative not to hide information or display it in untruthful ways.
Much recent visualization research explores how this imperative trades off against considerations like perceptual accuracy and affordances (e.g., ~\cite{Bertini2020-notallscatterplots, borkin2013memorableViz, padilla2016-binscalar}).
Our work approaches these issues by focusing on disclosure tactics and how visualizations can contain a disclosure gap. 
\revision{We distinguish disclosure tactics from emphasis tactics, which have received more attention for their impact on visualization interpretation \protect\cite{correll2020-yaxis, fygenson2024-arrangemarks, xiong2022-iconarray, long2024-cutYaxis}} \revision{and are therefore not the focus of our analysis.} 

Other theories relevant to disclosure characterize the process by which a chart is constructed in terms of a series of transforms. 
The first chronologically is Card et al.'s Information Visualization Reference Model~\cite{Card1999}, which distinguishes (i) the transforms that happen between a data source and a table and (ii) the transforms that happen between a table and a visual abstraction. 
Although multiple theories extend or adapt this model (e.g.,~\cite{Vickers2013-CategoryTheory, kindlmann2014-AVD, Wu2024-transforms}), two are of particular relevance to our work.
Kindlmann and Scheidegger's Algebraic Visualization Design (AVD)~\cite{kindlmann2014-AVD} introduces the idea that changes in data $\alpha$ should correspond to similarly important-seeming changes in a visualization $\omega$, and vice-versa.
They identify different ways that a visualization can be ill-formed or illegible, dubbed \textit{AVD vulnerabilities} (see Section \ref{qualitativeAnalysis}), based on the non-correspondence between input and output transforms.
Similar to others who have used these vulnerabilities in conceptual analyses to identify strengths and weaknesses in visualization design~\cite{correll2019-LooksGoodToMe, correll2023-terubozu, McNutt2020-Mirages}, we adopt AVD vulnerabilities as a framework for understanding the design implications of disclosure tactics.

Recently, Wu and Chang~\cite{Wu2024-transforms} extend Card et al.'s model in ways that help to formalize the problem of disclosure in visualization.
They differentiate (i) visual encoding transforms between a table and a visual abstraction from (ii) \textit{``design-specific transforms''} that preprocess an original data table in ways needed to generate a particular visualization.
These design-specific transforms are what we call disclosure tactics.
Whereas Wu and Chang develop formal notation around this distinction and critique the way these transforms are often conflated with encoding choices in visualization APIs, our work focuses on providing a systematic description of the disclosure tactics available to designers.

%% file: 3_methods.tex
\section{Method}

We conducted a qualitative analysis to identify the different disclosure tactics in visualization techniques. 
For our analysis, we developed a large corpus of diverse data visualization techniques included in research papers published at top visualization conferences and journals. 
Our ability to distinguish between disclosure tactics often depends on our knowledge of the original data,
which by definition \revision{is} not obvious in a visualization with a disclosure gap. 
Therefore, for our analysis, we used descriptions of original data provided in the text of academic articles or the implementation details included in supplemental materials to verify our classification of disclosure tactics. 
We made minimal assumptions about data representation and transformations beyond what was explicitly stated by the authors.
This is 
why our coding approach requires a corpus of scholarly articles about visualization techniques and not a mere corpus of example visualizations.
Through this analysis, we systematically identified and characterized the disclosure choices made in these visualizations, providing a framework for understanding how these decisions shape the information presented to viewers. 

\subsection{Building a Corpus}
We built our corpus in two phases as shown in Figure \ref{fig:flowchart}. 
In phase 1, we focused on gathering a broad range of data visualization techniques from papers that described them in sufficient detail. To this effect, we collected all the full papers published at IEEE VIS from 1990 - 2024. 
We then selected papers based on search terms including ``experiment'', ``evaluation'', ``technique'', or ``empirical'' in the title, abstract, or keywords section, since we needed access to the implementation details of the data visualizations and assumed such papers would document them. 
We limited our corpus to 2D, non-interactive visualizations representing tabular data to maintain a manageable scope.

After initial rounds of qualitative coding, we observed that many interesting disclosure gaps were present in visualizations that focused on uncertainty, distributions, visualization literacy, and AVD. In phase 2, we explored this further by collecting unique examples of visualization techniques published at venues such as ACM CHI, TVCG, CG\&A, and EG/CGF, focusing on search terms such as uncertainty and literacy. We used inclusion criteria similar to Phase 1.   
Our final corpus included 425 examples from 132 research papers. 
While our corpus is descriptive of the visualization techniques in the literature, it is not representative of the frequency with which disclosure tactics are applied to visualizations.  

\begin{figure*}[ht]
  \centering
  \includegraphics[width=\textwidth]
{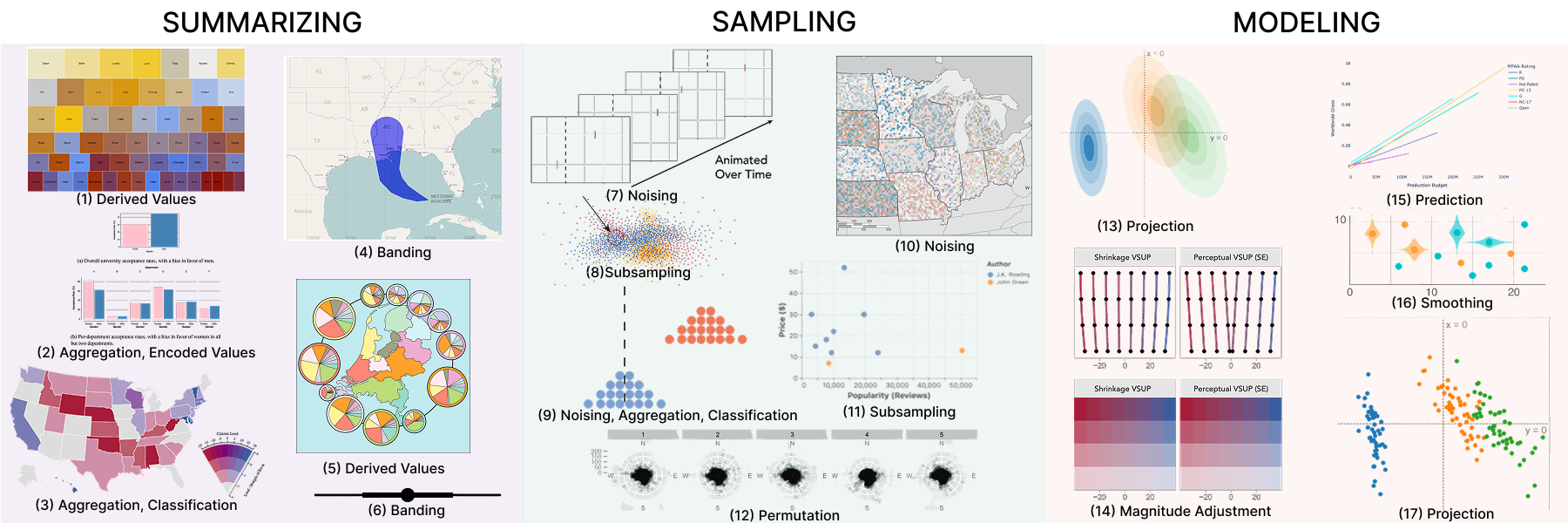}
  \setlength{\abovecaptionskip}{-3mm}
  \setlength{\belowcaptionskip}{-6mm}
  \caption{Examples of visualization techniques from our corpus that we analyzed and classified into the three families of disclosure tactics: \textcolor{summary}{summarizing} (examples from ~\cite{correll2018-VSUPs, correll2017-blackhat, Kay2024-ggdist, liu2018-visualizingCyclone, speckmann2010necklace, sondag2017stableTreemap}), \textcolor{sampling}{sampling} (examples from~\cite{wickham2010-graphicalLineups, kale2020visualStrategies, gaba2022comparison, hagh2007weaving, chen2014visualScatterplot}), and \textcolor{functional}{modeling} (examples from~\cite{kay2019-VSUP, sarma2022evaluatingMissingValue, gortler2019-uncertaintyPCA, kruchten2023metrics}).}
  \label{fig:disclosureList}
\end{figure*}

Some of the visualizations we analyzed are shown in Figure \ref{fig:disclosureList}. 
See \href{https://github.com/krisha-mehta/DisclosureInDataVis}{Supplemental Material} for (1) a spreadsheet of all the papers we surveyed, (2) the selection and refinement of papers, and (3) a codebook containing the full analysis of our corpus with screenshots of examples.

\subsection{Qualitative Analysis}
\label{qualitativeAnalysis}
We started our analysis by open-coding visualizations in the corpus,
describing possible disclosure gaps with
the datasets used to create them. 
This set the stage for subsequent analysis, enabling us to identify patterns in how design choices in visualizations impact disclosure.
\looseness=-1

Next, we used deductive coding~\cite{miles1994-qualitativeAnalysis} to place our analysis in dialogue with 
\revision{the framework of}
algebraic visualization design (AVD) \cite{kindlmann2014-AVD}. 
AVD enabled us to describe potential ambiguities in visualization interpretation (dubbed AVD vulnerabilities \cite{kindlmann2014-AVD}) that arise when transformations in data and visualizations do not correspond clearly. Based on their potential shortcomings, we classified each visualization in our corpus into four vulnerability types defined by AVD, as detailed below.
\begin{enumerate}[nolistsep]
    \item \textbf{Hallucinators} \revision{are superficial changes in the data} substantially alter the appearance of the visualization. For example, adding jitter to a dotplot \cite{few2017-datavisJitter} can create patterns in the visualizations that are absent in the dataset.
    \item \textbf{Confusers} \revision{are meaningful changes in the}
    data \revision{that} are
    impossible to detect in
    \revision{the}
    visualization. For example, \revision{Anscombe's quartet of datasets \cite{anscombe1973graphs} are identical if we plot only summary statistics, so swapping between them is a confuser}. \revision{Section 4.3 distinguishes confusers caused by disclosure tactics from those we dub \textit{perceptual confusers}.}
    \item \textbf{Jumblers} \revision{are meaningful changes in the data for which the resulting visualization change is perceptible but not readly interpretable}. For example, a normalized choropleth map can show changes in the original data, however, isolating the specific factors contributing to those changes can be difficult \cite{ge2024-vframer}.
    \item \textbf{Misleaders} \revision{are} important-seeming changes in the visualization \revision{that} do not correspond to similarly important changes in the original data. For example, a histogram with too many bins can highlight artifacts or patterns that are not truly significant \cite{correll2023-terubozu}.
\end{enumerate}AVD helps us analyze the visualization design space in terms of how choices about information disclosure correspond to choices about what information is rendered illegible or uninterpretable in a visualization.

During our qualitative analysis, along with disclosure tactics, a different set of design tactics emerged. We dub them \textit{emphasis tactics}, choices that do not impact disclosure but influence the potential interpretation of a chart. 
These include choices about scales (e.g., \cite{correll2020-yaxis, long2024-cutYaxis, witt2019graph}) and layouts (e.g., \cite{fygenson2024-arrangemarks, xiong2022-iconarray}).
We leave the analysis of emphasis tactics to future work, as they are out the of scope of our investigation.   

After multiple iterations of open coding and discussion,
\revision{our research team}
developed 13 distinct inductive codes ~\cite{miles1994-qualitativeAnalysis} describing operations that affect information disclosure in every visualization in the corpus. We verified our codes against the implementation details available for each visualization and made revisions where necessary. We were able to verify 125 examples, of which 5 initially were classified incorrectly by our team. 
We further categorized these operations into high-level families of disclosure tactics using an affinity diagramming procedure~\cite{miles1994-qualitativeAnalysis}.
\looseness=-1

%% file: 4_results.tex
\section{Results}
\label{results}

Through our qualitative analysis, we characterize how disclosure gaps are made in a diverse sample of visualization techniques. 
Our analysis produces a descriptive taxonomy of \textit{disclosure tactics}, data operations that are used to distort or hide information from the original data. 
We differentiate these tactics based on their consequences for disclosure, providing descriptions of data transformations at the level of their purpose rather than their exact implementation.

\subsection{Disclosure Tactics}
\label{disclosureTactics}
Our analysis surfaces 13 operations that impact information disclosure in visualizations, which we group into three \textit{families of disclosure tactics}.
Every visualization in our corpus employs at least one tactic from these families, which we describe in more detail below. 

\textcolor{summary}{\textbf{Summarizing}} (N = 314) changes the schema of the data representation for the purposes of grouping or partitioning data, or changing the units of analysis. This often entails computing summary statistics as a stand-in for the original data, which in effect changes the data representation to a summary. However, some operators (\eg{} classification, categorization) merely change how variables are expressed without themselves computing summary statistics. 
The following codes describe the disclosure tactics that perform \textcolor{summary}{summarizing}:
\begin{enumerate}[nolistsep]
    \item \textbf{Classification} \includesvg[scale=0.90]{images/Classification1.svg} involves partitioning continuous or ordinal data into discrete groups based on specific criteria, \eg{} grouping people based on their height.  
    Classification can make individual values contained within each partition indistinguishable, instead highlighting the features of the partition.
    \item \textbf{Categorization} \includesvg{images/Categorization.svg} 
    combines nominal values into groups that are often predefined, \eg{} grouping counties into states. 
    The use of this tactic enables designers to reduce the cardinality of nominal variables and focus on the categories may prefer to visualize. 
    \item \textbf{Aggregation} \includesvg[scale=0.85]{images/Aggregation.svg} collapses data values into summary statistics using an aggregate function, \eg{} calculating the mean temperature. The use of aggregation changes the data representation to a single summary statistic, or one statistic per data partition.
    \item \textbf{Banding} \includesvg{images/Threshold.svg} involves choosing a set of cut points that determines ranges of values to be visualized or boundaries on a distribution. It helps focus on specific regions of a distribution or highlight values above or below a given benchmark. Values that fall outside the cut points are often omitted or shown using different markings. Unlike classification results in non-overlapping partitions, banding can produce nested partitions.  
    \item \textbf{Derived Values} \includesvg[scale=0.8]{images/Derived_Values.svg} combines two or more variables of a dataset into one, \eg{} calculating the credit score of a person. 
    Importantly, derived values can obscure the details of the component variables. 
    \item \textbf{Encoded Values} \includesvg{images/Encode.svg} chooses specific columns of the dataset.
    One such example shown in Figure \ref{fig:disclosureList}.2 is Simpson's paradox, where the apparent pattern changes depending on what variables are encoded. 
    Importantly, non-encoding of columns is the silent default for most grammar of graphics implementations~\cite{wilkinson2012grammar}.
\end{enumerate}

\textcolor{sampling}{\textbf{Sampling}} (N = 163) is used to visualize a finite set of discrete points. The codes from our analysis that perform \textcolor{sampling}{sampling} include:
\begin{enumerate}[nolistsep]
    \item \textbf{Subsampling} \includesvg[scale=0.8]{images/SubSampling.svg} creates a subset of a sample-based dataset. The representativeness of a subsample depends on the sampling algorithm, the number of subsamples, and whether the subsample was drawn with or without replacement. 
    \item \textbf{Noising} \includesvg[scale=0.8]{images/Noising.svg} assumes a data generating model to produce samples, \eg{} showing hypothetical outcomes drawn from a normal distribution based on observed values with measurement error. We see noising used to show what an underlying phenomenon may look like based on known summary statistics, and to mask sensitive data in charts such as differentially private scatterplots \cite{panavas2023investigatingDiffPrivate, Zhang2021_DiffPrivacy}.
    
    \item \textbf{Permutation} \includesvg[scale=0.85]{images/Permutation.svg} changes the arrangement of values in a specific column to break association with other variables. We see this tactic used in lineup protocols of scatterplots where either the x or y values are shuffled to create a ``null model'' for the relation between the two variables \cite{wickham2010-graphicalLineups}.
\end{enumerate}

\textcolor{functional}{\textbf{Modeling}} (N = 79) involves using an algorithm to interpolate over infinitely small increments of the data domain, thus changing the data representation to the input and output of a function. The following codes from our analysis use \textcolor{functional}{modeling}:
\begin{enumerate}[nolistsep]
    \item \textbf{Smoothing} \includesvg[scale=0.8]{images/Smoothing.svg} fits a smooth curve on data points to show the inferred shape of the underlying distribution. For example, density plots and gradient plots use this tactic. 
    \item \textbf{Magnitude adjustment} \includesvg[scale=0.8]{images/Magnitude_adjustment.svg} modifies data values based on an adjustment function. We see this tactic used to correct for known inaccuracies in end-user perception or in data itself~\cite{kay2019-VSUP,yang2023-SubjectiveProb}.
    \item \textbf{Prediction} \includesvg{images/Modeling.svg} involves using a model to represent the relation between different variables and using that model to generate expected values for one or more variables in the original dataset, \eg{} regression analysis. 
    \item \textbf{Projection} \includesvg[scale=0.9]{images/Projection.svg} involves mapping a dataset from its original variables into an embedding space, \eg{} using PCA for dimensionality reduction. This serves to create an abstract data representation with a coordinate space corresponding to model internals. 
\end{enumerate}

Additionally, we require another disclosure tactic in our taxonomy for cases when there are no modifications to the original data. 
We call this tactic \textbf{full disclosure}, which entails visualizing everything in the original data representation. 
A dot plot where every point on the plot corresponds to a data point in a one-column dataset~\cite{correll2023-terubozu} is an example of such a tactic. 
Because the number of distinct variables we can effectively encode in a chart is limited, full disclosure only supports small or simple datasets, making it infeasible for most real-world data. 

\subsection{How Disclosure Tactics Are Used}
Families of disclosure tactics offer different \textbf{levels of detail} about the underlying dataset. 
For instance, \textcolor{sampling}{sample-based} charts offer information only about discrete points present in a sample, but they describe these points with high specificity. 
\textcolor{summary}{Summary-based} charts show the data at a far more coarse granularity with each group or partition described by a single summary statistic. 
\textcolor{functional}{Model-based} charts extrapolate patterns across the entire data domain that are possible to observe in theory, and they describe infinitely fine loci on the data domain with computed values.
In Sections \ref{section5} and \ref{discussion}, we consider how different levels of detail may be required for various tasks, in different applications, or for different user populations based on their graphical or statistical literacy.

Most disclosure tactics entail some \textbf{flexibility in implementation}, often in the form of a choice of algorithm or tuning of parameters. 
For example, all \textcolor{sampling}{sampling} tactics require the designer to choose a sampling algorithm.
\textcolor{summary}{Summarizing} tactics such as classification \includesvg[scale=0.72]{images/Classification1.svg} and banding \includesvg[scale=0.72]{images/Threshold.svg} require the designer to choose parameters such as bin edges or cut points. 
All \textcolor{functional}{modeling} tactics entail a choice of functional form or model, many of these with a particular set of parameters. 
In Section \ref{section5}, we show how this flexibility in implementation can be used to fine-tune disclosure gaps for a particular task, and how a lack of flexibility to reveal a particular signal can make a disclosure tactic misleading or undesirable in certain situations.

Sometimes we see disclosure tactics used together in \textbf{composition patterns}, which describe how multiple tactics can be used in tandem to render a visualization with a disclosure gap.
Three key composition patterns emerged from our analysis:
\begin{enumerate}[nolistsep]
    \item \textbf{Branch} creates separate copies of the dataset, where different disclosure tactics can be applied to each copy. These multiple copies can then be used to create separate layers of an image or frames of an animation, \eg{} in hypothetical outcome plots.
    \item \textbf{Chain} applies disclosure tactics to the data in a sequential order, where the resulting disclosure gap tends to be sensitive to this order. This enables us to describe multi-step data processing in terms of our operators from Section \ref{disclosureTactics}.
    \item \textbf{Combine} describes how two or more data representations can integrated into a visualization. This includes both joining or concatenating data tables and using different data representations to render different layers of the same visualization.
\end{enumerate}
Analyzing composition patterns revealed that some tactics, such as classification \includesvg[scale=0.72]{images/Classification1.svg}, categorization \includesvg[scale=0.72]{images/Categorization.svg}, and derived values \includesvg[scale=0.72]{images/Derived_Values.svg}, are seldom used in isolation. 
Instead, they are frequently 
used
with tactics such as aggregation \includesvg[scale=0.72]{images/Aggregation.svg} and encoded values \includesvg[scale=0.72]{images/Encode.svg} to create a disclosure gap.
For example, consider how creating a histogram requires a chaining pattern of classification \includesvg[scale=0.72]{images/Classification1.svg}  $\rightarrow$ aggregation \includesvg[scale=0.72]{images/Aggregation.svg}, and that classification \includesvg[scale=0.72]{images/Classification1.svg} alone would not produce a summary representation.
Sections \ref{section4.3} and \ref{section5} consider examples that use multiple tactics in composition patterns.

\begin{figure*}[t]
  \centering
  \includesvg[width=\textwidth, height = 8.25cm]{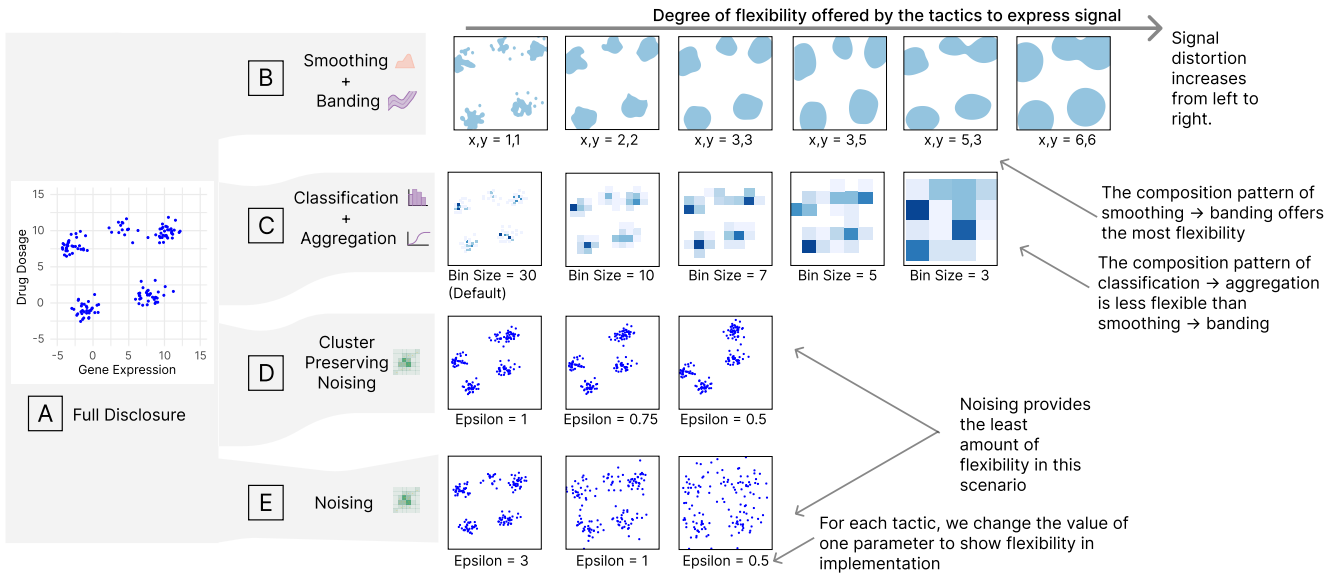}
  \setlength{\abovecaptionskip}{-3mm}
  \setlength{\belowcaptionskip}{-6mm}
  \caption{The different 
  disclosure tactics Fred the biostatistician applies to create visualizations that hide individual points while showing clusters.}
  \label{fig:Fred}
\end{figure*}

\subsection{How Disclosure Tactics Influence AVD Violations}
\label{section4.3}
\revision{We set out to investigate how disclosure tactics help explain a chart's susceptibility to certain AVD vulnerabilities\protect\cite{kindlmann2014-AVD},} \revision{especially the information loss that occurs due to the use disclosure tactics.}
By identifying AVD issues across our corpus, we find that  \textit{confusers} and \textit{jumblers} are often caused directly by disclosure tactics. 
In contrast, \textit{hallucinators} and \textit{misleaders} typically arise due to concomitant emphasis choices that are sometimes forced by the adoption of certain disclosure tactics. 
This distinction between vulnerabilities directly caused by disclosure tactics vs. those that stem from necessary design decisions made downstream helps to clarify the origin, and thus the nature, of AVD issues. 

\textit{Confusers} were the most frequent AVD vulnerabilities in our corpus (N = 307) due to the fact that they tend to be a direct consequence of disclosure tactics. 
The specific causes of confusers can vary depending on the tactic. 
The most common cause of confusers in our corpus was the omission of variables due to the use of the encoded values tactic \includesvg[scale=0.72]{images/Encode.svg}, which is a type of disclosure gap present in most visualizations. 
More generally, we see that confusers often result from the way families of disclosure tactics transform the distributional representation of the data. 
\textcolor{summary}{Summary-based} tactics like classification \includesvg[scale=0.72]{images/Classification1.svg} and banding \includesvg[scale=0.72]{images/Threshold.svg} can hide changes in the original data by reducing granularity. 
On the other hand, the \textcolor{sampling}{sampling} family of disclosure tactics can cause confusers by presenting imprecise or unrepresentative values for individual points. 
By obscuring fine-grained details,
\textcolor{functional}{modeling} tactics
can make it difficult to detect subtle patterns in the original data representation. 

However, we distinguish a subset of confusers that are not caused directly by disclosure tactics. Certain confusers (N = 12), which we dub \textit{perceptual confusers}, arise because emphasis tactics such as scaling choices render attributes of the underlying dataset imperceptible, such that meaningful variations appear in the visualization at less than one JND. Our analysis did not identify a systematic relationship between perceptual confusers and choices about disclosure tactics.

\textit{Jumblers}, though less frequent in our corpus (N = 73), occur when there is a lack of transparency about how a disclosure tactic is implemented.
For example, in visualizations created using derived values \includesvg[scale=0.72]{images/Derived_Values.svg}, jumblers arise when changes in the derived values cannot be traced back to changes in the source columns. 
All visualizations 
created using magnitude adjustment \includesvg[scale=0.72]{images/Magnitude_adjustment.svg} are jumblers, as the 
adjustment function is opaque and may be hard to decode even with a descriptive legend.
\looseness=-1

In contrast, hallucinators and misleaders have more indirect relationships with disclosure tactics, usually stemming instead from 
emphasis tactics.
Although emphasis tactics are out of scope for our analysis, we recognize that their use tends to be concomitant to choices about disclosure because disclosure tactics determine the data representation encoded in a visualization, thus limiting the set of mark types, scales, and layout choices that would result in a well-formed visualization. 
To understand how adopting a disclosure tactic filters the visualization design space, consider dotplots as an example. The designer applies a classification tactic \includesvg[scale=0.72]{images/Classification1.svg} to bin an initially \textcolor{sampling}{sample-based} representation, and now she must avoid overplotting. Unless she wants to switch to a \textcolor{summary}{summary} representation (e.g., a bar chart of counts), she needs to make a concomitant emphasis choice of either stacking, jittering, or varying the opacity of the dots. 
Thus, 
using a particular disclosure tactic can force a concomitant emphasis choice to preserve legibility.

This is the kind of situation where disclosure tactics have a role in creating hallucinators and misleaders.
All \textit{hallucinators} in our corpus (N = 30) encode a \textcolor{sampling}{sample-based} data representation, meaning they either use disclosure tactics from the \textcolor{sampling}{sampling} family or the designer initially accesses data in a \textcolor{sampling}{sample-based} format. 
%
Hallucinators arise
because \textcolor{sampling}{sample-based} data representations necessitate choices about layout such as sort order---\eg{} in a scatterplot, an icon array, or in frames an animated hypothetical outcome plot---changes which lead to circumstantial patterns in the visualization that do not reflect something intrinsic to the data.
\textit{Misleaders} in our analysis (N = 14) have a less consistent relationship with disclosure tactics. 
A potentially misleading non-correspondence between data and visualization can be introduced directly by a disclosure tactic (\eg{} in a histogram with too many bins~\cite{correll2023-terubozu}), but more often misleaders reflect concomitant layout choices (\eg{} jittering on a scale representing a continuous variable) 
As such, disclosure tactics live upstream of hallucinators and misleaders in the rendering process but tend not to directly determine their presence.

%% file: 5_implications.tex
\section{Making Choices about Disclosure in Visualization}
\label{section5}

\begin{figure*}[t]
  \centering
  \includesvg[width=\textwidth]{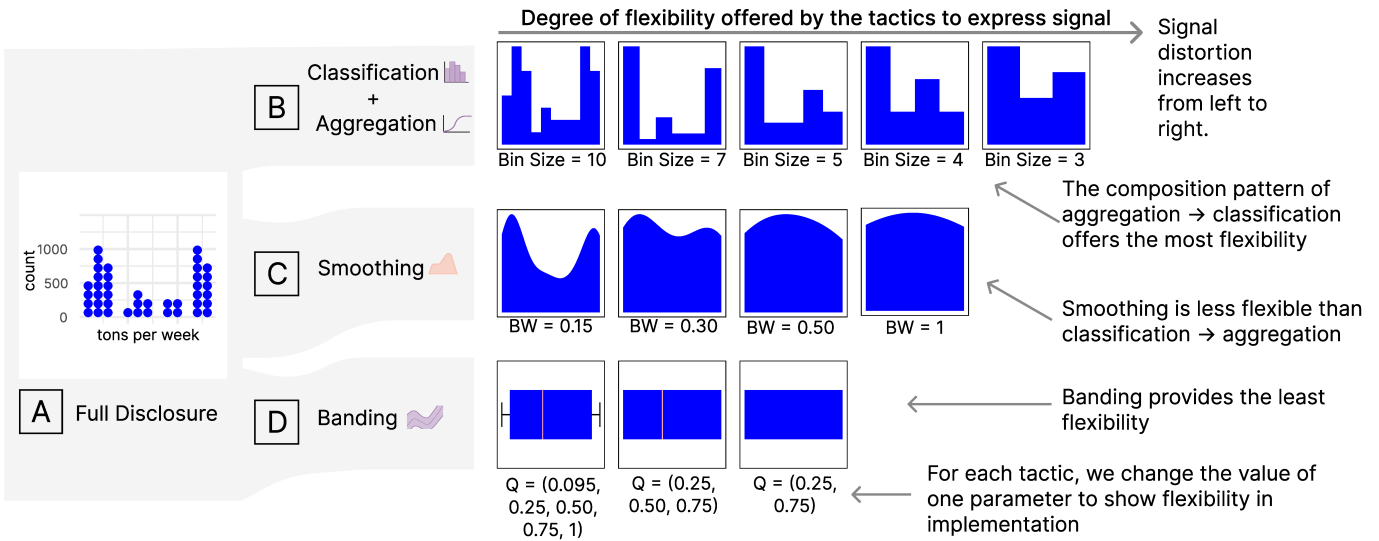}
  \setlength{\abovecaptionskip}{-3.5mm}
  \setlength{\belowcaptionskip}{-6mm}
  \caption{The different tactics Anne applies to create visualizations that can persuade her partner organizations to sign the NDA.}
  \label{fig:Anne}
\end{figure*}

Disclosure tactics provide a rich vocabulary for understanding how design choices impact the potential informativeness of visualizations.
In particular, our taxonomy gives a systematic account of the various ways that specific \textit{signals} (i.e., task-relevant pieces of information) can be revealed, distorted, or hidden by a visualization.
This expands on a particularly important but under-theorized implication of algebraic visualization design (AVD)~\cite{kindlmann2014-AVD}, namely, that most visualizations are confusers that hide some input signal. 
Whereas AVD suggests that designers must ``choose your confuser'', typically by a de facto method of guess-and-check, our taxonomy enables a more proactive approach we call ``tune your confuser'', whereby a visualization designer can reason systematically about how confusers are made in practice in order to achieve a desired disclosure gap. 
\revision{We assume that designers bring situated knowledge about their data and users, which help them reason about what signals are important to show or hide in a given situation.}\looseness=-1

The \textbf{key benefits of our taxonomy for visualization designers} are: (1) disclosure tactics and their composition patterns provide a way to ideate alternative techniques for achieving design goals related to disclosure; and (2) most disclosure tactics deal in degrees of signal distortion, affording designers \textit{flexibility} to distort some signals without distorting others.
The first benefit arises from the fact that disclosure tactics are composable operators with distinct purposes; the second results from the fact that each disclosure tactic entails a limited, and thus often predictable, set of options for fine-tuning disclosure gaps via algorithms and parameters used to implement them.
In order to demonstrate these benefits, we consider a series of design scenarios where visualization designers who are newly deputized with the vocabulary of disclosure address distinct trade-offs around signal distortions. 

\subsection{Flexible Privacy Protection}
\label{section5.1}

Reconsider the example of Fred (Section \ref{intro}), a bio-statistician at a university medical center who works with sensitive health data.
Figure \ref{fig:teaser} presents a scenario where Fred receives a request for data access from a pharmaceutical company to aid in drug discovery. 
\textit{The company wants to identify patient cohorts that cluster together in terms of expression of a particular gene and the effective dose of an immunosuppressant.}

Fred's goal is to give a straightforward answer to the pharmaceutical company's query while respecting privacy regulations. 
On the one hand, he wants to show clusters of gene expression and effective dose as clearly as possible, ideally avoiding an unnecessarily complex approach that might be difficult to explain.
On the other hand, he must avoid revealing information specific to individuals in the dataset, so they cannot be reidentified.
In essence, Fred's dilemma (the canonical problem of individual privacy) involves a trade-off around hiding one signal while minimally distorting another. 

Drawing on our vocabulary of disclosure, Fred considers alternative visualizations for showing clusters on gene expression and effective dose, and we compare these options with full disclosure in a scatterplot (Figure \ref{fig:Fred}A).
Reasoning that a tactic from the \textcolor{summary}{summarizing} family might work well to group individuals together (analogous to $k$-anonymity~\cite{sweeney2002-kanon}), Fred adopts a classification \includesvg[scale=0.72]{images/Classification1.svg} $\rightarrow$ aggregation \includesvg[scale=0.72]{images/Aggregation.svg} pattern to generate a heatmap.
However, as Figure \ref{fig:Fred}C shows, an out-of-the-box classification \includesvg[scale=0.72]{images/Classification1.svg} approach is not flexible enough to reveal clusters without drawing bins so fine that they isolate individual records, violating Fred's privacy objective. This is particularly challenging due to the sparseness of this dataset and the closeness of clusters.
Fred could instead rely on a non-uniform binning procedure that properly implements $k$-anonymity~\cite{dasgupta2013measuringPrivacyPreserve} while aligning bin edges with cluster boundaries. 
However, this would result in an uneven-looking heatmap that could be difficult to explain to the company.
This demonstrates how classification \includesvg[scale=0.72]{images/Classification1.svg} can be insufficiently flexible for revealing task-relevant patterns such as clusters unless we are willing to implement and explain bespoke approaches to preserve these signals (e.g.,
~\cite{jenks1967-naturalBreaks}).

Next, Fred considers a different solution where, rather than using \textcolor{summary}{summarization} to hide individual values completely, he tries using tactics from the \textcolor{sampling}{sampling} family to distort the individual records in the dataset. 
This leads him to try creating a differentially private (DP) scatterplot \cite{panavas2023investigatingDiffPrivate}, which relies on the noising tactic \includesvg[scale=0.72]{images/Noising.svg} to distort values of gene expression and effective dose. 
However, Fred recognizes that the success of this approach will depend on flexibility in the design of the sampling algorithm.
He reasons that a naive approach to DP (Figure \ref{fig:Fred}E) will fail out-of-the box to preserve the clusters in cases where it adds a sufficient amount of noise to obscure individual values.
Similarly, Fred recognizes that even if he adopts a cluster-preserving DP algorithm (Figure \ref{fig:Fred}D), it's possible this won't necessarily preserve all signals that might be relevant to the company, such as the location and extent of each cluster.
By definition, \textcolor{sampling}{sampling} tactics like the noising algorithm \includesvg[scale=0.72]{images/Noising.svg} used in any implementation of DP will inevitably put Fred in the position of having to explain that some signals in the resulting visualization could be spurious.
This demonstrates how the effective use of flexibility in disclosure tactics, regardless of family, requires the designer to think deeply about how they conceptualize distinct signals in their use case.
We argue \revision{that} this demand for precision about what counts as signal is a benefit of a 
framework that prioritizes disclosure.

Finally, Fred realizes that his problem calls for the affordances of multiple families of tactics, \textcolor{functional}{modeling} to capture the clusters similar to a bespoke version of DP, but also \textcolor{summary}{summarization} to hide the individual values similar to what he intended with the heatmap.
Consequently, Fred decides to use a composition pattern of smoothing \includesvg[scale=0.72]{images/Smoothing.svg} $\rightarrow$ banding \includesvg[scale=0.72]{images/Threshold.svg} to generate a contour plot as a way of sharing clusters with the 
company.
Figure \ref{fig:Fred}B shows how this can easily produce a variety of images that clearly delineate clusters without revealing individual records.
Interestingly, Fred prefers an approach with a large disclosure gap, which \textcolor{summary}{summarizes} the clusters with a single 85\% containment interval on the underlying smoothed \includesvg[scale=0.72]{images/Smoothing.svg} density, rather than revealing superfluous concentric regions of density
(not shown).
This visualization is fine-tuned to reveal only the clusters with as simple a representation as possible, and unlike other approaches Fred considered, the contour plot does this without requiring a bespoke implementation.
This solution works 
because the \textcolor{functional}{modeling} tactic underlying the intermediate smoothing \includesvg[scale=0.72]{images/Smoothing.svg} preserves shape while glossing over low-level details. 
Additionally, the banding \includesvg[scale=0.72]{images/Threshold.svg} tactic can eliminate non-essential details such as ambiguous cluster edges and uneven density within clusters, supporting Fred's goal of a representation that requires little to no explanation.
\looseness=-1

\begin{figure*}
  \centering
  \includesvg[width=\textwidth]{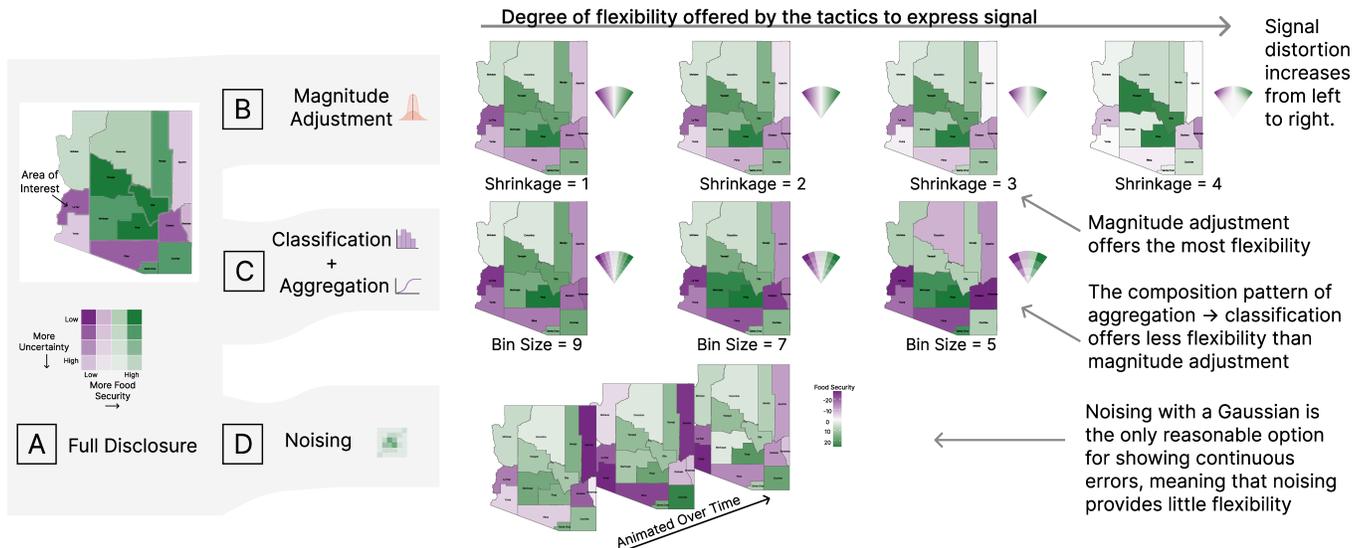}
  \setlength{\abovecaptionskip}{-4mm}
  \setlength{\belowcaptionskip}{-6.75mm}
  \caption{The different tactics Claudia applies to her geospatial data to create maps that show both food insecurity and data uncertainty.}
  \label{fig:Claudia}
\end{figure*}

\subsection{Trustworthy Data Sharing}
\label{section52}

Consider another case, the example of Anne, a data scientist in the supply chain industry whose job includes sharing data with partner organizations.
Anne works with hard-to-source proprietary data about fluctuations in the supply of raw materials for manufacturing, and as such her employer requires partner organizations to sign a non-disclosure agreement (NDA) as a part of their data sharing contract.
\textit{Her counterpart from a potential partner organization asks Anne if she has information about the supply of a particular mineral, specifically how frequently the amount produced exceeds 1000 tons per week, and they are hoping to gain a reliable way to anticipate how the availability of this resource could impact their business.}


Anne's goal is to be trusted as an information provider.
On the one hand, she wants to show visualizations that persuade her counterpart that she has data that can help them enough to justify signing an NDA.
On the other hand, she wants to distort the relative frequency of production greater than 1000 tons per week enough that she doesn't reveal too much detail about her employer's proprietary data in advance of a signed data sharing contract.
In essence, Anne's scenario involves a trade-off about how much to distort a specific signal in order to support both her counterpart's objective and her own.\footnote{A national security analyst would face similar problems summarizing data for personnel with different levels of security clearance.}


Anne considers a handful of alternative visualizations, and we compare how these options distort task-relevant signals relative to full disclosure of the supply dataset in a dotplot (Figure \ref{fig:Anne}A). 
First, Anne errs on the side of caution, reasoning that a tactic from the \textcolor{summary}{summarizing} family is likely her best bet for not revealing too much about her employer's data.
Anne considers sending her counterpart a boxplot (Figure \ref{fig:Anne}D), using a banding tactic \includesvg[scale=0.72]{images/Threshold.svg} to show only selected quantiles of the distribution. 
However, she recognizes that banding \includesvg[scale=0.72]{images/Threshold.svg} has little flexibility to reveal the skewed and lumpy shape of the distribution, leaving important signals for judging cumulative probability to her counterpart's imagination. 
This demonstrates that the least flexible tactics from the \textcolor{summary}{summarizing} family (i.e., banding, aggregation, derived values) tend to risk hiding a distribution's shape.

Next, recognizing that tactics from the \textcolor{functional}{modeling} family might be appropriate for revealing the shape of her dataset while still hiding low-level details, Anne considers sending her counterpart a probability density plot, adopting a smoothing tactic \includesvg[scale=0.72]{images/Smoothing.svg}. 
Figure \ref{fig:Anne}C shows how depending on the parameterization of the smoothing tactic, this approach tends to distort peaks and troughs in the underlying distribution with limited flexibility not to consequently distort the relative frequency of production over 1000 tons per week.
This demonstrates a common difficulty with tactics in the \textcolor{functional}{modeling} family, similar to Fred's challenge fine-tuning a DP algorithm (see Section \ref{section5.1}), that preserving any signals pertaining to detailed structure in the data requires more sophisticated models than are often available out-of-the-box.

Finally, Anne reasons that she needs a \textcolor{summary}{summarizing} approach, similar to what she initially considered with the boxplot (Figure \ref{fig:Anne}D), but one that affords greater flexibility to reveal low-level details.
She decides to share a histogram, opting for a classification \includesvg[scale=0.72]{images/Classification1.svg} $\rightarrow$ aggregation \includesvg[scale=0.72]{images/Aggregation.svg} composition pattern.
Figure \ref{fig:Anne}B shows that this approach affords her the flexibility to choose a number of bins that better preserves relative frequency in the distribution. 
Additionally, classification \includesvg[scale=0.72]{images/Classification1.svg} in particular enables Anne to fine-tune the visualization such that her counterpart's benchmark of interest (supply greater than 1000 tons per week) falls within the middle of a bin, distorting a precise answer to their query in advance of signing an NDA.
In this sense, out-of-the-box approaches to classification \includesvg[scale=0.72]{images/Classification1.svg} are more flexible for conveying cumulative probability than out-of-the-box approaches to smoothing \includesvg[scale=0.72]{images/Smoothing.svg}, and thus histograms are a logically better choice than densities for Anne's disclosure problem.
This demonstrates how our taxonomy of disclosure tactics can help to fine-tune the distortion of a particular signal to address design trade-offs around trustworthy communication.

\subsection{Advocacy with Geospatial Uncertainty}
\label{section53}
Consider the case of Claudia, an analyst employed at a non-profit who helps lobby her state government.
Imagine Claudia has been tasked with identifying counties for which to advocate the deployment of social services to alleviate food insecurity. 
Unfortunately, the available data on food insecurity is highly uncertain in some places, and to qualify for services, a county's need must be considered both acute and statistically reliable.
Additionally, because most members of the state legislature lack statistical training, Claudia should avoid unnecessary complexity in communicating about uncertainty.
\textit{To help identify eligible counties without loss of geospatial context, Claudia's boss asks for some kind of simple choropleth map showing ``hot spots'' for food insecurity.}
\looseness=-1

Claudia's goal is to provide a visualization that helps to persuade the state legislature that an action is justified.
This means she needs to show where action is warranted by clearly communicating both food insecurity and uncertainty. 
Only the areas with both acute and statistically reliable needs should stand out. 
In order to avoid making her visualization overly complex, Claudia believes that she should avoid showing a low level of detail.\footnote{Note that the issue of complexity in visualization can also be understood in ways that are not strictly about disclosure~\cite{windhager2024complexity}.}
In essence, Claudia's problem is to reveal two signals in tandem, central tendency and reliability, without producing a visualization that is overwhelming in its level of detail. 

Claudia considers choropleth map variants that use different uncertainty representations for identifying counties with acute and reliable food insecurity. 
We compare these options with full disclosure in a map with a bivariate color palette~\cite{trumbo1981-theoryBivariate} 
(Figure \ref{fig:Claudia}A), an approach that tends to communicate poorly because color and opacity are perceptually non-separable~\cite{correll2018-VSUPs}, an important consideration outside the scope of our taxonomy. 
To achieve her goal of revealing both task-relevant signals, food insecurity and uncertainty, Claudia could use a noising tactic \includesvg[scale=0.72]{images/Noising.svg} to render uncertainty representations such as HOPs~\cite{hullman2015-hop} (Figure \ref{fig:Claudia}D) or dithered choropleth maps (not shown).
However, she recognizes that approaches in the \textcolor{sampling}{sampling} family inherently reveal more low-level detail than she wants to present to the city council.

Based on her need for a simple representation, Claudia gravitates toward a \textcolor{summary}{summarizing} approach.
Specifically, she considers Value Suppressing Uncertainty Palettes (VSUPs) \cite{correll2018-VSUPs}, which use a bespoke composition pattern of hierarchical classification \includesvg[scale=0.72]{images/Classification1.svg} $\rightarrow$ aggregation \includesvg[scale=0.72]{images/Aggregation.svg}.
Claudia knows that the composition pattern in VSUPs will distort both signals, food insecurity and uncertainty, and she considers whether this approach enables her to fine-tune the level of distortion to meet her needs (Figure \ref{fig:Claudia}C).
Unfortunately, the \textcolor{summary}{summarizing} tactics used to make VSUPs result in either too few color categories to distinguish hot spots or a legend with too high a level of detail. 
This demonstrates how \textcolor{summary}{summarizing} tactics in general trade off level of detail with signal distortion.
\looseness=-1

Recognizing that her use case calls for greater flexibility, Claudia opts for tactics from the \textcolor{functional}{modeling} family.
To make this change, she pivots to an alternative version of VSUPs~\cite{kay2019-VSUP}, which uses a magnitude adjustment tactic \includesvg[scale=0.72]{images/Magnitude_adjustment.svg} rather than \textcolor{summary}{summarization} to highlight only the reliable hot spots.
Figure \ref{fig:Claudia}B shows that this approach affords flexibility to tune the degree of magnitude adjustment \includesvg[scale=0.72]{images/Magnitude_adjustment.svg} so that the uncertainty signal does not interfere with the acute need signal or vice versa.
Although any visualization created through magnitude adjustment \includesvg[scale=0.72]{images/Magnitude_adjustment.svg} is a jumbler in the parlance of AVD \cite{kindlmann2014-AVD}---a chart where changes in the image cannot be clearly attributed to specific changes in the data---adopting the framework of disclosure tactics to reason about what signals are distorted provides some reassurance that magnitude adjustment VSUPs \includesvg[scale=0.72]{images/Magnitude_adjustment.svg} are appropriate in Claudia's use case.
This example demonstrates that sometimes we need bespoke \textcolor{functional}{modeling} tactics to pre-compute task-relevant signals, especially in cases where alternative approaches would hide or distort these signals in undesirable ways.
\looseness=-1

%% file: 6_discussion.tex
\section{Discussion}
\label{discussion}
Our work highlights the importance of information disclosure as a lens for understanding data visualization. 
Thinking of design problems in terms of disclosure leads to a focus on \textit{what signals} are revealed, distorted, or hidden in a visualization and \textit{exactly how}. 
This approach enables designers to address common trade-offs in data visualization, as we demonstrate in Section \ref{section5} through a series of scenarios around privacy, data sharing, and persuasion.
Although these scenarios each involve different trade-offs, 
they all share a \textbf{core tension between the virtues of comprehensibility and transparency}.
Comprehensibility often requires a designer to use some form of data reduction to facilitate ease of understanding and avoid overwhelming end-users, especially when working with large or complex datasets.
In contrast, transparency requires a designer to avoid disclosure gaps, at least when they omit information that's relevant to end-users.
Navigating this tension requires specificity about possible solutions (i.e., visualizations with distinct approaches to disclosure) and how we evaluate them (i.e., what signal distortions are admissible), a degree of specificity eschewed by existing guidelines and principles for visualization.
Our taxonomy of disclosure tactics provides a  helpful conceptual framework for understanding these issues, such that we can anticipate and explain what visualizations are useful for specific design goals around disclosure.

The core contribution of our investigation is an analysis of \textit{disclosure tactics}, data operations that determine what information is revealed or hidden in a visualization.
We distinguish these operations from \textit{emphasis tactics}, choices about mark types, scales, and layouts that influence how the information encoded in a chart might be subsequently interpreted, in order to isolate issues of disclosure as much as possible.
Section \ref{results} provides a descriptive account of disclosure tactics, grouping these operations into \textbf{three families} based on the kinds of data representations they produce and their affordances:
\begin{itemize}[nolistsep]
    \item \textcolor{summary}{\textbf{Summarizing:}} Tactics in this family tend to reduce the level of detail about a dataset through grouping, partitioning, and changing metrics. All of the least flexible tactics live in this family, making \textcolor{summary}{summarizing} tactics most useful in situations where available groups or partitions have a clear relationship to task-relevant signals (e.g., Section \ref{section52}) but less useful in situations where task-relevant signals are hard to operationalize (e.g., Section \ref{section5.1}) or hard to isolate (e.g., Section \ref{section53}).
    \item \textcolor{sampling}{\textbf{Sampling:}} Tactics in this family facilitate showing individual data points, a relatively low of level detail. \textcolor{sampling}{Sampling} tactics are as flexible as the algorithm that implements them, whether samples are produced by subsampling observed data or by noising with a model. As such sampling tactics are most helpful when the algorithm can be fine-tuned to capture task-relevant signals (e.g., Section \ref{section5.1}) but less useful in situations where point-level signals might be spurious (e.g., Section \ref{section5.1}) or overwhelmingly complex for the audience (e.g., Section \ref{section53}).
    \item \textcolor{functional}{\textbf{Modeling:}} Tactics in this family show datasets by capturing their relationships through abstraction, providing a mapping between variables at a potentially infinite level of detail within the data domain. Similar to sampling, \textcolor{functional}{modeling} tactics can be made extremely flexible to the extent that their implementation captures task-relevant signals (e.g., Section \ref{section53}). Although simple models can be well-matched to a task (e.g., Section \ref{section5.1}), they tend to distort signals they are not designed to preserve (e.g., Section \ref{section52}). \looseness=-1
\end{itemize}
These families of disclosure tactics can be useful for design ideation in part because they facilitate mostly non-overlapping sets of visualization techniques, meaning that they provide a way to conceptualize what kind of visualization a situation calls for.
However, in order to fine-tune a visualization for a specific design goal around disclosure or explain why a particular technique does or doesn't work, we usually need to think about individual tactics and how much flexibility we have in implementing them (as we demonstrate in Section \ref{section5}). 

Disclosure tactics provide an \textbf{explanatory framework} for understanding potential failure modes of visualization, supplementing algebraic visualization design (AVD) principles~\cite{kindlmann2014-AVD}.
Whereas AVD analysis tends to entail \textit{ex post facto} dissection of examples to categorize issues of legibility in visualizations (e.g.,~\cite{kindlmann2014-AVD, correll2019-LooksGoodToMe, McNutt2020-Mirages, correll2023-terubozu}), disclosure tactics make AVD work more proactively to help designers generate and provisionally evaluate solutions to problems (e.g., see Section \ref{section5}).
Consider how AVD accounts for disclosure gaps with categories of vulnerabilities called \textit{confusers} (i.e., omission of information) and \textit{jumblers} (i.e., non-transparent mapping from data to visual channels).
Even if a designer knows what these terms mean, they might be hard pressed to use AVD to choose among possible vulnerabilities or to explain exactly why these issues arise.
In contrast, disclosure tactics describe how confusers and jumblers are made and provide a vocabulary for deciding---without a method of guess-and-check---if and when they might pose problems for end-users.
Our work also leads to new conceptualizations of AVD vulnerabilities, such as the distinction between \textit{disclosure confusers} (i.e., actual omission) vs. \textit{perceptual confusers} (i.e., invisibility due to scaling and layout), and the fact that \textit{hallucinators} (i.e., non-data-encoding visual patterns) are often created by layout choices forced by the use of \textcolor{sampling}{sample-based} data representations (see Section \ref{section4.3}).
We argue that these advancements set the stage for \textit{future theory development}, which might proceed by formalizing the operators and composition patterns that emerged from our descriptive analysis.
Future work should consider how our taxonomy of disclosure tactics, with its relationship to AVD~\cite{kindlmann2014-AVD}, can be synthesized with other theoretical frameworks for visualization, such as statistical decision theory~\cite{wu2024-rationalAgentBenchmark} and cognitive accounts of visualization interpretation~\cite{Padilla2018-framework}.

\subsection{Visualization Research Through the Lens of Disclosure} 
We argue that adopting disclosure as a framework can be transformative for visualization research, by offering a change in perspective on problems of deep importance to the field.

\textbf{Authoring Tools.}
Our work invites us to envision building visualization authoring tools in ways that make disclosure a first-order consideration. 
Recent work critiques visualization APIs (e.g., VegaLite \cite{satyanarayan2016-vegalite}) for mixing declaration of design-specific data transforms into the specification of visual encodings \cite{Wu2024-transforms}, and our work echoes this critique. 
\revision{Instead, we suggest that visualization tools focused on disclosure start with declarations of which signals are relevant and what the author's disclosure goals are for each signal, i.e., whether to reveal, distort, or hide them.}
This would enable the interactive selection of disclosure tactics and composition patterns that might be suitable for the author's goals, followed by choices about emphasis tactics such as mark types, scales, and layouts, 
\revision{thereby pruning the design space 
of
possible visualizations.}
\revision{This approach scaffolds authoring around the sequence of data processing steps, making AVD vulnerabilities}~\cite{kindlmann2014-AVD} \revision{more detectable by highlighting how early 
use of
disclosure tactics may introduce confusers and jumblers, before
subsequent
emphasis tactics risk creating hallucinators and misleaders.}
\revision{This process transparency aligns with recent calls to highlight ``seams'' in visualization design\protect\cite{wood2018design, Hengesbach2022_UndoingSeam}.}
\revision{Our taxonomy might serve as the basis for a set of abstractions for directly declaring disclosure tactics, or for recommending possible implementations of higher-level design goals about disclosure, thereby helping practitioners apply the framework.} To enable such systems, future work should investigate how visualization authors prefer to declare design goals related to disclosure.

\textbf{Visualization Literacy.}
The notion of disclosure gaps offers a new way of conceptualizing challenges around visualization literacy.
Prior work on visualization literacy emphasizes the pitfalls of decoding visualizations with built-in preprocessing steps (e.g., normalization~\cite{ge2024-vframer, correll2016-surprisebayesian, McNutt2020-Mirages}, aggregation~\cite{McNutt2020-Mirages, Ge2023-Calvi, guo2017-WhatYouSeeSimpsonsParadox}), and the importance of learning graphical conventions to help recognize these pitfalls and avoid them during interpretation \cite{borner2019-literacy, adar2021_communicativeVis}. 
In complementary fashion, our taxonomy of disclosure tactics provides a systematic description of how the construction of visualizations can create such pitfalls, distinguishing disclosure gaps that distort or hide information from other issues of legibility \cite{kindlmann2014-AVD, McNutt2020-Mirages}.
As such, our framework may be useful in the design of items for visualization literacy assessments \eg \cite{lee2016vlat, boy2014principledVizLiteracy, cui2023-adaptiveLiteracy}. 
Additionally, we posit that knowledge of disclosure tactics may be helpful for \textit{reasoning backwards} from a visualization to possible datasets, thus offering a way to help end-users \textit{interpret visualizations defensively by identifying possible disclosure gaps}.
Future work should investigate whether understanding disclosure tactics aids in visualization interpretation, exploring possibilities around new educational materials and disclosure-based learning objectives.
Future work should also build automated interpretation aids that infer possible disclosure gaps from an input visualization, so assistive technologies can warn end-users about what a chart could never show about the underlying dataset.

\textbf{Uncertainty Visualization.}
We argue that disclosure gaps are an especially important consideration for uncertainty visualization.
Our analysis shows that creating uncertainty visualizations tends to involve the use of disclosure tactics to translate the original data into the distributional form required to render a particular technique.
For example, discrete encodings of uncertainty require \textcolor{sampling}{sampling} tactics unless the original representation is sample-based.
In fact, the distributional forms required to produce the various visual uncertainty representations offered by the probabilistic grammar of graphics\cite{pu2020-PGoG} and \texttt{ggdist}~\cite{Kay2024-ggdist} correspond to our three families of disclosure tactics.
This suggests that often designers will be forced to use disclosure tactics when visualizing uncertainty, whether they recognize it or not, and that consequent signal distortions or omissions likely play an outsized role in effective uncertainty communication (e.g., Section 5.3).
Similarly, for model interpretability techniques, which often summarize model outputs using uncertainty visualizations, disclosure can provide a helpful framework for formalizing what signals from a model are revealed, distorted, or hidden.
Future work should develop ways of ensuring that uncertainty visualizations and model interpretability techniques preserve task-relevant signals consistent with their design goals, rather than allowing tacit signal distortions and omissions to go unnoticed or uncorrected.
\looseness=-1

\textbf{Personalization and Accessibility.}
We argue that disclosure tactics can help to inform the way we personalize information displays or translate them to non-visual media in order to broaden access.
These potential benefits stem from the fact that disclosure tactics operationalize control over the level of detail in a data representation.
Prior work highlights the importance of controlling information granularity in visualizations to avoid cognitive overload~\cite{wu2023-data_data}, to facilitate different tasks~\cite{oscar2017-personal_viz,elavsky2022-chartability, Wu2024-transforms}, to make visualizations responsive to different sizes of display~\cite{kim2023-dupo}, when designing for older adults~\cite{price2016effects, galesic2009iconarray, nurgalieva2019-info_need_older_adults}, and for accessibility~\cite{kim2021accessible, sharif2023-uncertain_screenreader}.
This suggests that the vocabulary of disclosure tactics may be useful both (1) for end-users to express their desired level of detail in a personalized display and (2) for generating alt-text describing the content of a visualization at a ``statistical and relational'' level \cite{lundgard2021accessible}.
Future work should investigate how ideas and language about disclosure can complement existing approaches to translation for personalization and accessibility \cite{lee2020-reachingBroaderAudience, jung2021-communicatingWithoutVisuals}, including mixed-initiative personalization using AI \cite{schetinger2023doom} and the development of information loss metrics for translating visualizations \cite{kim-TaskOrientedInsights}. 

\textbf{Ethics of Disclosure.}
Although our work emphasizes more prosocial uses of disclosure tactics (e.g., in Section 5), we must take seriously the potential for visualization designers to use disclosure tactics in ways that are subtly and intentionally misleading.
For example, recent work on algorithmic persuasion describes how selective information disclosure can be used to induce desired beliefs in an audience \cite{dughmi2016algorithmicPersuasion, gao2022-inferenceSelectiveDisclosure}, which in theory is instrumental to manipulating behavior \cite{kamenica2011bayesian, kamenica2019bayesian}. 
Our taxonomy of disclosure tactics extends existing conceptualizations of rhetorical visualization~\cite{hullman2011-rhetoric, markant2023-viz_persuade, markant2022-attitude_change_minds, McNutt2020-Mirages, correll2017-blackhat, pandey2014-persuasive, karduni2021-bayesianCog} in ways that might enable bad actors but which also might offer a framework for developing enforcement mechanisms for ethical standards.
We envision that our vocabulary for disclosure can be used to craft: (1) auditing checklists to help detect deceptive visualization; (2) policies prohibiting specific tactics, or the hiding of specific signals, in settings such as courtrooms; and (3) assistive technologies that attempt to warn audiences about signals a visualization would be expected to conceal by design.
Future work should engage stakeholders in various application domains to co-design ethical standards and investigate how to help visualization non-specialists recognize and reason about possible disclosure gaps.

\subsection{Limitations}
Our analysis was limited to 2D, non-interactive visualizations of tabular data.
We made this scoping decision to keep our corpus of examples to a reasonable size and to focus an already-broad investigation.
As such, our results do not reflect disclosure tactics that might be unique to working with other abstract data types such as trees, graphs, and networks, unstructured data such as images and text, or 3D rendering.
However, we speculate that these kinds of data will often be transformed into a tabular format prior to visualizing, and that our taxonomy can be easily extended to account for any data transformations that might be truly unique to working with non-tabular data formats.
In contrast, although we do not analyze interactive visualizations, we believe our results have implications for them, specifically that disclosure tactics might be helpful for reasoning about informational differences between states of an interactive visualization that the user might navigate to.
Future work should extend and adapt our taxonomy to these settings.

Although our analysis relies on our interpretation of the disclosure tactics that might have been used to generate a given example, we went to great lengths to verify our coding process where possible.
Specifically, we sourced examples from academic articles, which we assumed would tend to include enough detail about how visualizations were created that we could check the disclosure tactics, and we used as an inclusion criterion the availability of supplemental materials that we deemed likely sufficient to verify our coding process.
We were able to verify our coding of disclosure tactics in 125/425 examples.
Examples that we could not verify were mostly due to the lack of 
visualization-generating code
in the supplemental material or links that are no longer valid. 
We argue this verification process makes our analysis much more rigorous than it would be otherwise, however, our approach should still be understood as interpretive in the sense that the disclosure tactics we present 
in part 
reflect 
our expert judgment of examples.

\revision{
By conceptualizing how disclosure gaps arise in data visualizations, this study highlights one facet of a larger problem.
Other factors, such as user interpretations and affect}~\cite{Lee-Robbins2023-affective}\revision{, remain central to visualization. Thus, }
although we discuss the practical design implications of our taxonomy of disclosure tactics (Section \ref{section5}), we caution that our conceptual analysis cannot speak to how visualizations will be interpreted by end-users. 
Interpretation fundamentally depends on both the perceptual and cognitive experience of the user and emphasis tactics such as mark type, scale, and layout choices, neither of which was in scope for our investigation. Other factors such as the rhetorical placement of labels may also influence interpretation \cite{hullman2011-rhetoric}. Although textual components like annotations and labels do not always affect disclosure, they can affect it in ways that were not examined in our study, which focused specifically on the fidelity of visually encoded data.
Rather than giving a capacious account of visualization interpretation, our work mostly speaks to the availability of information required to perform various tasks.
We think of this as placing a rough upper bound on performance, in the sense that certain disclosure gaps can distort or hide task-relevant signals in ways that would logically preclude effective communication.
Otherwise, we intentionally avoid speculating about end-user interpretations beyond referencing the findings of prior work whose research was designed to speak to these issues.
Future empirical evaluations with human subjects are required to address issues of interpretation and to provide a fuller account of how disclosure relates to visualization effectiveness.

\section{Conclusion}
We contribute a taxonomy of disclosure tactics providing a systematic description of various ways that visualizations can reveal, distort, or hide information from the viewer, which we derive from a content analysis of 425 examples of 2D visualization techniques for tabular data sampled from the visualization literature.
Our results group disclosure tactics into three families of data operations that impact disclosure by producing different kinds of data representations: \textcolor{summary}{summary}, \textcolor{sampling}{sample}, and \textcolor{functional}{model}.
We find that these disclosure tactics have predictable consequences for violations of algebraic visualization design principles~\cite{kindlmann2014-AVD}, and we provide a series of examples that demonstrate how our taxonomy provides a useful framework for reasoning through common design trade-offs around disclosure.
Our work offers a new way of conceptualizing important problems for visualization research, such as authoring, literacy, uncertainty, personalization, and ethical design.

%% file: main.bbl
\begin{thebibliography}{10}

\bibitem{adar2021_communicativeVis}
E.~Adar and E.~Lee.
\newblock Communicative visualizations as a learning problem.
\newblock {\em IEEE Trans. Visual Comput. Graphics}, 27(2):946--956, 2021. \href{https://doi.org/10.1109/TVCG.2020.3030375}
{doi: {{%
10\hspace{.1pt}\discretionary{.}{%
}{.}\hspace{.4pt}1109\discretionary{/}{%
}{/}TVCG\hspace{.1pt}\discretionary{.}{%
}{.}\hspace{.4pt}2020\hspace{.1pt}\discretionary{.}{%
}{.}\hspace{.4pt}3030375}}}


\bibitem{anscombe1973graphs}
F.~J. Anscombe.
\newblock Graphs in statistical analysis.
\newblock {\em The American Statistician}, 27(1):17--21, 1973. \href{https://doi.org/10.2307/2682899}
{doi: {{%
10\hspace{.1pt}\discretionary{.}{%
}{.}\hspace{.4pt}2307\discretionary{/}{%
}{/}2682899}}}


\bibitem{bertin1983-semiology}
J.~Bertin.
\newblock {\em Semiology of graphics}.
\newblock University of Wisconsin Press, 1983.

\bibitem{Bertini2020-notallscatterplots}
E.~Bertini, M.~Correll, and S.~Franconeri.
\newblock Why shouldn’t all charts be scatter plots? beyond precision-driven visualizations.
\newblock In {\em IEEE VIS}, pp. 206--210, 2020. \href{https://doi.org/10.1109/VIS47514.2020.00048}
{doi: {{%
10\hspace{.1pt}\discretionary{.}{%
}{.}\hspace{.4pt}1109\discretionary{/}{%
}{/}VIS47514\hspace{.1pt}\discretionary{.}{%
}{.}\hspace{.4pt}2020\hspace{.1pt}\discretionary{.}{%
}{.}\hspace{.4pt}00048}}}


\bibitem{borkin2013memorableViz}
M.~A. Borkin, A.~A. Vo, Z.~Bylinskii, P.~Isola, S.~Sunkavalli, A.~Oliva, and H.~Pfister.
\newblock What makes a visualization memorable?
\newblock {\em IEEE Trans. Visual Comput. Graphics}, 19(12):2306--2315, 2013. \href{https://doi.org/10.1109/TVCG.2013.234}
{doi: {{%
10\hspace{.1pt}\discretionary{.}{%
}{.}\hspace{.4pt}1109\discretionary{/}{%
}{/}TVCG\hspace{.1pt}\discretionary{.}{%
}{.}\hspace{.4pt}2013\hspace{.1pt}\discretionary{.}{%
}{.}\hspace{.4pt}234}}}


\bibitem{boy2014principledVizLiteracy}
J.~Boy, R.~A. Rensink, E.~Bertini, and J.-D. Fekete.
\newblock A principled way of assessing visualization literacy.
\newblock {\em IEEE Trans. Visual Comput. Graphics}, 20(12):1963--1972, 2014. \href{https://doi.org/10.1109/TVCG.2014.2346984}
{doi: {{%
10\hspace{.1pt}\discretionary{.}{%
}{.}\hspace{.4pt}1109\discretionary{/}{%
}{/}TVCG\hspace{.1pt}\discretionary{.}{%
}{.}\hspace{.4pt}2014\hspace{.1pt}\discretionary{.}{%
}{.}\hspace{.4pt}2346984}}}


\bibitem{borner2019-literacy}
K.~Börner, A.~Bueckle, and M.~Ginda.
\newblock Data visualization literacy: Definitions, conceptual frameworks, exercises, and assessments.
\newblock {\em PNAS}, 116(6):1857--1864, 2019. \href{https://doi.org/10.1073/pnas.1807180116}
{doi: {{%
10\hspace{.1pt}\discretionary{.}{%
}{.}\hspace{.4pt}1073\discretionary{/}{%
}{/}pnas\hspace{.1pt}\discretionary{.}{%
}{.}\hspace{.4pt}1807180116}}}


\bibitem{Card1999}
S.~K. Card, J.~Mackinlay, and B.~Shneiderman.
\newblock {\em Readings in information visualization: using vision to think.}
\newblock Morgan Kaufmann, 1999.

\bibitem{chen2014visualScatterplot}
H.~Chen, W.~Chen, H.~Mei, Z.~Liu, K.~Zhou, W.~Chen, W.~Gu, and K.-L. Ma.
\newblock Visual abstraction and exploration of multi-class scatterplots.
\newblock {\em IEEE Trans. Visual Comput. Graphics}, 20(12):1683--1692, 2014. \href{https://doi.org/10.1109/TVCG.2014.2346594}
{doi: {{%
10\hspace{.1pt}\discretionary{.}{%
}{.}\hspace{.4pt}1109\discretionary{/}{%
}{/}TVCG\hspace{.1pt}\discretionary{.}{%
}{.}\hspace{.4pt}2014\hspace{.1pt}\discretionary{.}{%
}{.}\hspace{.4pt}2346594}}}


\bibitem{correll2023-terubozu}
M.~Correll.
\newblock Teru teru b{\=o}zu: Defensive raincloud plots.
\newblock In {\em Comput. Graphics Forum}, vol.~42, pp. 235--246. Wiley Online Library, 2023. \href{https://doi.org/10.1111/cgf.14826}
{doi: {{%
10\hspace{.1pt}\discretionary{.}{%
}{.}\hspace{.4pt}1111\discretionary{/}{%
}{/}cgf\hspace{.1pt}\discretionary{.}{%
}{.}\hspace{.4pt}14826}}}


\bibitem{correll2020-yaxis}
M.~Correll, E.~Bertini, and S.~Franconeri.
\newblock Truncating the y-axis: Threat or menace?
\newblock In {\em Proc. CHI},  12 pages, p. 1–12. ACM, New York, 2020. \href{https://doi.org/10.1145/3313831.3376222}
{doi: {{%
10\hspace{.1pt}\discretionary{.}{%
}{.}\hspace{.4pt}1145\discretionary{/}{%
}{/}3313831\hspace{.1pt}\discretionary{.}{%
}{.}\hspace{.4pt}3376222}}}


\bibitem{correll2017-blackhat}
M.~Correll and J.~Heer.
\newblock Black hat visualization.
\newblock In {\em Workshop on Dealing with Cognitive Biases in Visualisations (DECISIVe), IEEE VIS}, vol.~1, p.~10, 2017.

\bibitem{correll2016-surprisebayesian}
M.~Correll and J.~Heer.
\newblock Surprise! bayesian weighting for de-biasing thematic maps.
\newblock {\em IEEE Trans. Visual Comput. Graphics}, 23(1):651--660, 2017. \href{https://doi.org/10.1109/TVCG.2016.2598618}
{doi: {{%
10\hspace{.1pt}\discretionary{.}{%
}{.}\hspace{.4pt}1109\discretionary{/}{%
}{/}TVCG\hspace{.1pt}\discretionary{.}{%
}{.}\hspace{.4pt}2016\hspace{.1pt}\discretionary{.}{%
}{.}\hspace{.4pt}2598618}}}


\bibitem{correll2019-LooksGoodToMe}
M.~Correll, M.~Li, G.~Kindlmann, and C.~Scheidegger.
\newblock Looks good to me: Visualizations as sanity checks.
\newblock {\em IEEE Trans. Visual Comput. Graphics}, 25(1):830--839, 2019. \href{https://doi.org/10.1109/TVCG.2018.2864907}
{doi: {{%
10\hspace{.1pt}\discretionary{.}{%
}{.}\hspace{.4pt}1109\discretionary{/}{%
}{/}TVCG\hspace{.1pt}\discretionary{.}{%
}{.}\hspace{.4pt}2018\hspace{.1pt}\discretionary{.}{%
}{.}\hspace{.4pt}2864907}}}


\bibitem{correll2018-VSUPs}
M.~Correll, D.~Moritz, and J.~Heer.
\newblock Value-suppressing uncertainty palettes.
\newblock In {\em Proc. CHI},  11 pages, p. 1–11. ACM, New York, 2018. \href{https://doi.org/10.1145/3173574.3174216}
{doi: {{%
10\hspace{.1pt}\discretionary{.}{%
}{.}\hspace{.4pt}1145\discretionary{/}{%
}{/}3173574\hspace{.1pt}\discretionary{.}{%
}{.}\hspace{.4pt}3174216}}}


\bibitem{cui2023-adaptiveLiteracy}
Y.~Cui, L.~W. Ge, Y.~Ding, F.~Yang, L.~Harrison, and M.~Kay.
\newblock Adaptive assessment of visualization literacy.
\newblock {\em IEEE Trans. Visual Comput. Graphics}, 30(1):628--637, 2024. \href{https://doi.org/10.1109/TVCG.2023.3327165}
{doi: {{%
10\hspace{.1pt}\discretionary{.}{%
}{.}\hspace{.4pt}1109\discretionary{/}{%
}{/}TVCG\hspace{.1pt}\discretionary{.}{%
}{.}\hspace{.4pt}2023\hspace{.1pt}\discretionary{.}{%
}{.}\hspace{.4pt}3327165}}}


\bibitem{dasgupta2013measuringPrivacyPreserve}
A.~Dasgupta, M.~Chen, and R.~Kosara.
\newblock Measuring privacy and utility in privacy-preserving visualization.
\newblock In {\em Comput. Graphics Forum}, vol.~32, pp. 35--47. Wiley Online Library, 2013. \href{https://doi.org/10.1111/cgf.12142}
{doi: {{%
10\hspace{.1pt}\discretionary{.}{%
}{.}\hspace{.4pt}1111\discretionary{/}{%
}{/}cgf\hspace{.1pt}\discretionary{.}{%
}{.}\hspace{.4pt}12142}}}


\bibitem{dughmi2016algorithmicPersuasion}
S.~Dughmi and H.~Xu.
\newblock Algorithmic bayesian persuasion.
\newblock In {\em Proc. STOC}, STOC '16,  14 pages, p. 412–425. ACM, New York, 2016. \href{https://doi.org/10.1145/2897518.2897583}
{doi: {{%
10\hspace{.1pt}\discretionary{.}{%
}{.}\hspace{.4pt}1145\discretionary{/}{%
}{/}2897518\hspace{.1pt}\discretionary{.}{%
}{.}\hspace{.4pt}2897583}}}


\bibitem{elavsky2022-chartability}
F.~Elavsky, C.~Bennett, and D.~Moritz.
\newblock How accessible is my visualization? evaluating visualization accessibility with chartability.
\newblock In {\em Comput. Graphics Forum}, vol.~41, pp. 57--70. Wiley Online Library, 2022. \href{https://doi.org/10.1111/cgf.14522}
{doi: {{%
10\hspace{.1pt}\discretionary{.}{%
}{.}\hspace{.4pt}1111\discretionary{/}{%
}{/}cgf\hspace{.1pt}\discretionary{.}{%
}{.}\hspace{.4pt}14522}}}


\bibitem{few2017-datavisJitter}
S.~Few and P.~Edge.
\newblock The datavis jitterbug: Let’s improve an old dance.
\newblock {\em Perceptual Edge, Visual Business Intelligence Newsletter, April/May/June}, 2017.

\bibitem{fotheringham1991modifiable}
A.~S. Fotheringham and D.~W.~S. Wong.
\newblock The modifiable areal unit problem in multivariate statistical analysis.
\newblock {\em Environ. Plann. A: Econ. Space}, 23(7):1025--1044, 1991. \href{https://doi.org/10.1068/a231025}
{doi: {{%
10\hspace{.1pt}\discretionary{.}{%
}{.}\hspace{.4pt}1068\discretionary{/}{%
}{/}a231025}}}


\bibitem{fygenson2024-arrangemarks}
R.~Fygenson, S.~Franconeri, and E.~Bertini.
\newblock The arrangement of marks impacts afforded messages: Ordering, partitioning, spacing, and coloring in bar charts.
\newblock {\em IEEE Trans. Visual Comput. Graphics}, 30(01):1008--1018, jan 2024. \href{https://doi.org/10.1109/TVCG.2023.3326590}
{doi: {{%
10\hspace{.1pt}\discretionary{.}{%
}{.}\hspace{.4pt}1109\discretionary{/}{%
}{/}TVCG\hspace{.1pt}\discretionary{.}{%
}{.}\hspace{.4pt}2023\hspace{.1pt}\discretionary{.}{%
}{.}\hspace{.4pt}3326590}}}


\bibitem{gaba2022comparison}
A.~Gaba, V.~Setlur, A.~Srinivasan, J.~Hoffswell, and C.~Xiong.
\newblock Comparison conundrum and the chamber of visualizations: An exploration of how language influences visual design.
\newblock {\em IEEE Trans. Visual Comput. Graphics}, 29(1):1211--1221, 2023. \href{https://doi.org/10.1109/TVCG.2022.3209456}
{doi: {{%
10\hspace{.1pt}\discretionary{.}{%
}{.}\hspace{.4pt}1109\discretionary{/}{%
}{/}TVCG\hspace{.1pt}\discretionary{.}{%
}{.}\hspace{.4pt}2022\hspace{.1pt}\discretionary{.}{%
}{.}\hspace{.4pt}3209456}}}


\bibitem{galesic2009iconarray}
M.~Galesic, R.~Garcia-Retamero, and G.~Gigerenzer.
\newblock Using icon arrays to communicate medical risks: overcoming low numeracy.
\newblock {\em Health psychology}, 28(2):210, 2009. \href{https://doi.org/10.1037/a0014474}
{doi: {{%
10\hspace{.1pt}\discretionary{.}{%
}{.}\hspace{.4pt}1037\discretionary{/}{%
}{/}a0014474}}}


\bibitem{gao2022-inferenceSelectiveDisclosure}
Y.~Gao.
\newblock Inference from selectively disclosed data.
\newblock {\em arXiv preprint arXiv:2204.07191}, 2022. \href{https://doi.org/10.48550/arXiv.2204.07191}
{doi: {{%
10\hspace{.1pt}\discretionary{.}{%
}{.}\hspace{.4pt}48550\discretionary{/}{%
}{/}arXiv\hspace{.1pt}\discretionary{.}{%
}{.}\hspace{.4pt}2204\hspace{.1pt}\discretionary{.}{%
}{.}\hspace{.4pt}07191}}}


\bibitem{Ge2023-Calvi}
L.~W. Ge, Y.~Cui, and M.~Kay.
\newblock Calvi: Critical thinking assessment for literacy in visualizations.
\newblock In {\em Proc. CHI},  article no. 815,  18 pages. ACM, New York, 2023. \href{https://doi.org/10.1145/3544548.3581406}
{doi: {{%
10\hspace{.1pt}\discretionary{.}{%
}{.}\hspace{.4pt}1145\discretionary{/}{%
}{/}3544548\hspace{.1pt}\discretionary{.}{%
}{.}\hspace{.4pt}3581406}}}


\bibitem{ge2024-vframer}
L.~W. Ge, M.~Easterday, M.~Kay, E.~Dimara, P.~Cheng, and S.~L. Franconeri.
\newblock V-framer: Visualization framework for mitigating reasoning errors in public policy.
\newblock In {\em Proc. CHI},  article no. 390,  15 pages. ACM, New York, 2024. \href{https://doi.org/10.1145/3613904.3642750}
{doi: {{%
10\hspace{.1pt}\discretionary{.}{%
}{.}\hspace{.4pt}1145\discretionary{/}{%
}{/}3613904\hspace{.1pt}\discretionary{.}{%
}{.}\hspace{.4pt}3642750}}}


\bibitem{guo2017-WhatYouSeeSimpsonsParadox}
Y.~Guo, C.~Binnig, and T.~Kraska.
\newblock What you see is not what you get! detecting simpson's paradoxes during data exploration.
\newblock In {\em Proceedings of the 2nd Workshop on Human-In-the-Loop Data Analytics}, HILDA '17,  article no. 2,  5 pages. ACM, New York, 2017. \href{https://doi.org/10.1145/3077257.3077266}
{doi: {{%
10\hspace{.1pt}\discretionary{.}{%
}{.}\hspace{.4pt}1145\discretionary{/}{%
}{/}3077257\hspace{.1pt}\discretionary{.}{%
}{.}\hspace{.4pt}3077266}}}


\bibitem{gortler2019-uncertaintyPCA}
J.~Görtler, T.~Spinner, D.~Streeb, D.~Weiskopf, and O.~Deussen.
\newblock Uncertainty-aware principal component analysis.
\newblock {\em IEEE Trans. Visual Comput. Graphics}, 26(1):822--831, 2020. \href{https://doi.org/10.1109/TVCG.2019.2934812}
{doi: {{%
10\hspace{.1pt}\discretionary{.}{%
}{.}\hspace{.4pt}1109\discretionary{/}{%
}{/}TVCG\hspace{.1pt}\discretionary{.}{%
}{.}\hspace{.4pt}2019\hspace{.1pt}\discretionary{.}{%
}{.}\hspace{.4pt}2934812}}}


\bibitem{hagh2007weaving}
H.~Hagh-Shenas, S.~Kim, V.~Interrante, and C.~Healey.
\newblock Weaving versus blending: a quantitative assessment of the information carrying capacities of two alternative methods for conveying multivariate data with color.
\newblock {\em IEEE Trans. Visual Comput. Graphics}, 13(6):1270--1277, 2007. \href{https://doi.org/10.1109/TVCG.2007.70623}
{doi: {{%
10\hspace{.1pt}\discretionary{.}{%
}{.}\hspace{.4pt}1109\discretionary{/}{%
}{/}TVCG\hspace{.1pt}\discretionary{.}{%
}{.}\hspace{.4pt}2007\hspace{.1pt}\discretionary{.}{%
}{.}\hspace{.4pt}70623}}}


\bibitem{Hengesbach2022_UndoingSeam}
N.~Hengesbach.
\newblock Undoing seamlessness: Exploring seams for critical visualization.
\newblock In {\em Extended Abstracts of the 2022 CHI Conference on Human Factors in Computing Systems}, CHI EA '22,  article no. 364,  7 pages. ACM, New York, 2022. \href{https://doi.org/10.1145/3491101.3519703}
{doi: {{%
10\hspace{.1pt}\discretionary{.}{%
}{.}\hspace{.4pt}1145\discretionary{/}{%
}{/}3491101\hspace{.1pt}\discretionary{.}{%
}{.}\hspace{.4pt}3519703}}}


\bibitem{hullman2011-rhetoric}
J.~Hullman and N.~Diakopoulos.
\newblock Visualization rhetoric: Framing effects in narrative visualization.
\newblock {\em IEEE Trans. Visual Comput. Graphics}, 17(12):2231--2240, 2011. \href{https://doi.org/10.1109/TVCG.2011.255}
{doi: {{%
10\hspace{.1pt}\discretionary{.}{%
}{.}\hspace{.4pt}1109\discretionary{/}{%
}{/}TVCG\hspace{.1pt}\discretionary{.}{%
}{.}\hspace{.4pt}2011\hspace{.1pt}\discretionary{.}{%
}{.}\hspace{.4pt}255}}}


\bibitem{hullman2015-hop}
J.~Hullman, P.~Resnick, and E.~Adar.
\newblock Hypothetical outcome plots outperform error bars and violin plots for inferences about reliability of variable ordering.
\newblock {\em PloS one}, 10(11):e0142444, 2015. \href{https://doi.org/10.1371/journal.pone.0142444}
{doi: {{%
10\hspace{.1pt}\discretionary{.}{%
}{.}\hspace{.4pt}1371\discretionary{/}{%
}{/}journal\hspace{.1pt}\discretionary{.}{%
}{.}\hspace{.4pt}pone\hspace{.1pt}\discretionary{.}{%
}{.}\hspace{.4pt}0142444}}}


\bibitem{jenks1967-naturalBreaks}
G.~F. Jenks.
\newblock The data model concept in statistical mapping.
\newblock {\em International Yearbook of Cartography}, 7:186--190, 1967.

\bibitem{jung2021-communicatingWithoutVisuals}
C.~Jung, S.~Mehta, A.~Kulkarni, Y.~Zhao, and Y.-S. Kim.
\newblock Communicating visualizations without visuals: Investigation of visualization alternative text for people with visual impairments.
\newblock {\em IEEE Trans. Visual Comput. Graphics}, 28(1):1095--1105, 2022. \href{https://doi.org/10.1109/TVCG.2021.3114846}
{doi: {{%
10\hspace{.1pt}\discretionary{.}{%
}{.}\hspace{.4pt}1109\discretionary{/}{%
}{/}TVCG\hspace{.1pt}\discretionary{.}{%
}{.}\hspace{.4pt}2021\hspace{.1pt}\discretionary{.}{%
}{.}\hspace{.4pt}3114846}}}


\bibitem{kale2020visualStrategies}
A.~Kale, M.~Kay, and J.~Hullman.
\newblock Visual reasoning strategies for effect size judgments and decisions.
\newblock {\em IEEE Trans. Visual Comput. Graphics}, 27(2):272--282, 2021. \href{https://doi.org/10.1109/TVCG.2020.3030335}
{doi: {{%
10\hspace{.1pt}\discretionary{.}{%
}{.}\hspace{.4pt}1109\discretionary{/}{%
}{/}TVCG\hspace{.1pt}\discretionary{.}{%
}{.}\hspace{.4pt}2020\hspace{.1pt}\discretionary{.}{%
}{.}\hspace{.4pt}3030335}}}


\bibitem{kamenica2019bayesian}
E.~Kamenica.
\newblock Bayesian persuasion and information design.
\newblock {\em Annual Review of Economics}, 11(1):249--272, 2019. \href{https://doi.org/10.1146/annurev-economics-080218-025739}
{doi: {{%
10\hspace{.1pt}\discretionary{.}{%
}{.}\hspace{.4pt}1146\discretionary{/}{%
}{/}annurev\discretionary{%
}{-}{-}economics\discretionary{%
}{-}{-}080218\discretionary{%
}{-}{-}025739}}}


\bibitem{kamenica2011bayesian}
E.~Kamenica and M.~Gentzkow.
\newblock Bayesian persuasion.
\newblock {\em American Economic Review}, 101(6):2590–2615, October 2011. \href{https://doi.org/10.1257/aer.101.6.2590}
{doi: {{%
10\hspace{.1pt}\discretionary{.}{%
}{.}\hspace{.4pt}1257\discretionary{/}{%
}{/}aer\hspace{.1pt}\discretionary{.}{%
}{.}\hspace{.4pt}101\hspace{.1pt}\discretionary{.}{%
}{.}\hspace{.4pt}6\hspace{.1pt}\discretionary{.}{%
}{.}\hspace{.4pt}2590}}}


\bibitem{karduni2021-bayesianCog}
A.~Karduni, D.~Markant, R.~Wesslen, and W.~Dou.
\newblock A bayesian cognition approach for belief updating of correlation judgement through uncertainty visualizations.
\newblock {\em IEEE Trans. Visual Comput. Graphics}, 27(2):978--988, 2021. \href{https://doi.org/10.1109/TVCG.2020.3029412}
{doi: {{%
10\hspace{.1pt}\discretionary{.}{%
}{.}\hspace{.4pt}1109\discretionary{/}{%
}{/}TVCG\hspace{.1pt}\discretionary{.}{%
}{.}\hspace{.4pt}2020\hspace{.1pt}\discretionary{.}{%
}{.}\hspace{.4pt}3029412}}}


\bibitem{kay2019-VSUP}
M.~Kay.
\newblock How much value should an uncertainty palette suppress if an uncertainty palette should suppress value? statistical and perceptual perspectives, Oct 2019. \href{https://doi.org/10.31219/osf.io/6xcnw}
{doi: {{%
10\hspace{.1pt}\discretionary{.}{%
}{.}\hspace{.4pt}31219\discretionary{/}{%
}{/}osf\hspace{.1pt}\discretionary{.}{%
}{.}\hspace{.4pt}io\discretionary{/}{%
}{/}6xcnw}}}


\bibitem{Kay2024-ggdist}
M.~Kay.
\newblock ggdist: Visualizations of distributions and uncertainty in the grammar of graphics.
\newblock {\em IEEE Trans. Visual Comput. Graphics}, 30(1):414--424, 2024. \href{https://doi.org/10.1109/TVCG.2023.3327195}
{doi: {{%
10\hspace{.1pt}\discretionary{.}{%
}{.}\hspace{.4pt}1109\discretionary{/}{%
}{/}TVCG\hspace{.1pt}\discretionary{.}{%
}{.}\hspace{.4pt}2023\hspace{.1pt}\discretionary{.}{%
}{.}\hspace{.4pt}3327195}}}


\bibitem{kim2023-dupo}
H.~Kim, R.~Rossi, J.~Hullman, and J.~Hoffswell.
\newblock Dupo: A mixed-initiative authoring tool for responsive visualization.
\newblock {\em IEEE Trans. Visual Comput. Graphics}, 30(1):934--943, 2024. \href{https://doi.org/10.1109/TVCG.2023.3326583}
{doi: {{%
10\hspace{.1pt}\discretionary{.}{%
}{.}\hspace{.4pt}1109\discretionary{/}{%
}{/}TVCG\hspace{.1pt}\discretionary{.}{%
}{.}\hspace{.4pt}2023\hspace{.1pt}\discretionary{.}{%
}{.}\hspace{.4pt}3326583}}}


\bibitem{kim-TaskOrientedInsights}
H.~Kim, R.~Rossi, A.~Sarma, D.~Moritz, and J.~Hullman.
\newblock An automated approach to reasoning about task-oriented insights in responsive visualization.
\newblock {\em IEEE Trans. Visual Comput. Graphics}, 28(1):129--139, 2022. \href{https://doi.org/10.1109/TVCG.2021.3114782}
{doi: {{%
10\hspace{.1pt}\discretionary{.}{%
}{.}\hspace{.4pt}1109\discretionary{/}{%
}{/}TVCG\hspace{.1pt}\discretionary{.}{%
}{.}\hspace{.4pt}2021\hspace{.1pt}\discretionary{.}{%
}{.}\hspace{.4pt}3114782}}}


\bibitem{kim2021accessible}
N.~W. Kim, S.~C. Joyner, A.~Riegelhuth, and Y.~Kim.
\newblock Accessible visualization: Design space, opportunities, and challenges.
\newblock In {\em Comput. graphics forum}, vol.~40, pp. 173--188. Wiley Online Library, 2021. \href{https://doi.org/10.1111/cgf.14298}
{doi: {{%
10\hspace{.1pt}\discretionary{.}{%
}{.}\hspace{.4pt}1111\discretionary{/}{%
}{/}cgf\hspace{.1pt}\discretionary{.}{%
}{.}\hspace{.4pt}14298}}}


\bibitem{kindlmann2014-AVD}
G.~Kindlmann and C.~Scheidegger.
\newblock An algebraic process for visualization design.
\newblock {\em IEEE Trans. Visual Comput. Graphics}, 20(12):2181--2190, 2014. \href{https://doi.org/10.1109/TVCG.2014.2346325}
{doi: {{%
10\hspace{.1pt}\discretionary{.}{%
}{.}\hspace{.4pt}1109\discretionary{/}{%
}{/}TVCG\hspace{.1pt}\discretionary{.}{%
}{.}\hspace{.4pt}2014\hspace{.1pt}\discretionary{.}{%
}{.}\hspace{.4pt}2346325}}}


\bibitem{kruchten2023metrics}
N.~Kruchten, A.~M. McNutt, and M.~J. McGuffin.
\newblock Metrics-based evaluation and comparison of visualization notations.
\newblock {\em IEEE Trans. Visual Comput. Graphics}, 30(1):425--435, 2024. \href{https://doi.org/10.1109/TVCG.2023.3326907}
{doi: {{%
10\hspace{.1pt}\discretionary{.}{%
}{.}\hspace{.4pt}1109\discretionary{/}{%
}{/}TVCG\hspace{.1pt}\discretionary{.}{%
}{.}\hspace{.4pt}2023\hspace{.1pt}\discretionary{.}{%
}{.}\hspace{.4pt}3326907}}}


\bibitem{lee2020-reachingBroaderAudience}
B.~Lee, E.~K. Choe, P.~Isenberg, K.~Marriott, and J.~Stasko.
\newblock Reaching broader audiences with data visualization.
\newblock {\em IEEE Comput. Graphics Appl.}, 40(2):82--90, 2020. \href{https://doi.org/10.1109/MCG.2020.2968244}
{doi: {{%
10\hspace{.1pt}\discretionary{.}{%
}{.}\hspace{.4pt}1109\discretionary{/}{%
}{/}MCG\hspace{.1pt}\discretionary{.}{%
}{.}\hspace{.4pt}2020\hspace{.1pt}\discretionary{.}{%
}{.}\hspace{.4pt}2968244}}}


\bibitem{lee2016vlat}
S.~Lee, S.-H. Kim, and B.~C. Kwon.
\newblock Vlat: Development of a visualization literacy assessment test.
\newblock {\em IEEE Trans. Visual Comput. Graphics}, 23(1):551--560, 2017. \href{https://doi.org/10.1109/TVCG.2016.2598920}
{doi: {{%
10\hspace{.1pt}\discretionary{.}{%
}{.}\hspace{.4pt}1109\discretionary{/}{%
}{/}TVCG\hspace{.1pt}\discretionary{.}{%
}{.}\hspace{.4pt}2016\hspace{.1pt}\discretionary{.}{%
}{.}\hspace{.4pt}2598920}}}


\bibitem{Lee-Robbins2023-affective}
E.~Lee-Robbins and E.~Adar.
\newblock Affective learning objectives for communicative visualizations.
\newblock {\em IEEE Trans. Visual Comput. Graphics}, 29(01):1--11, jan 2023. \href{https://doi.org/10.1109/TVCG.2022.3209500}
{doi: {{%
10\hspace{.1pt}\discretionary{.}{%
}{.}\hspace{.4pt}1109\discretionary{/}{%
}{/}TVCG\hspace{.1pt}\discretionary{.}{%
}{.}\hspace{.4pt}2022\hspace{.1pt}\discretionary{.}{%
}{.}\hspace{.4pt}3209500}}}


\bibitem{liu2018-visualizingCyclone}
L.~Liu, L.~Padilla, S.~H. Creem-Regehr, and D.~H. House.
\newblock Visualizing uncertain tropical cyclone predictions using representative samples from ensembles of forecast tracks.
\newblock {\em IEEE Trans. Visual Comput. Graphics}, 25(1):882--891, 2019. \href{https://doi.org/10.1109/TVCG.2018.2865193}
{doi: {{%
10\hspace{.1pt}\discretionary{.}{%
}{.}\hspace{.4pt}1109\discretionary{/}{%
}{/}TVCG\hspace{.1pt}\discretionary{.}{%
}{.}\hspace{.4pt}2018\hspace{.1pt}\discretionary{.}{%
}{.}\hspace{.4pt}2865193}}}


\bibitem{long2024-cutYaxis}
S.~Long and M.~Kay.
\newblock To cut or not to cut? a systematic exploration of y-axis truncation.
\newblock In {\em Proc. CHI}, CHI '24,  article no. 207,  12 pages. ACM, New York, 2024. \href{https://doi.org/10.1145/3613904.3642102}
{doi: {{%
10\hspace{.1pt}\discretionary{.}{%
}{.}\hspace{.4pt}1145\discretionary{/}{%
}{/}3613904\hspace{.1pt}\discretionary{.}{%
}{.}\hspace{.4pt}3642102}}}


\bibitem{lundgard2021accessible}
A.~Lundgard and A.~Satyanarayan.
\newblock Accessible visualization via natural language descriptions: A four-level model of semantic content.
\newblock {\em IEEE Trans. Visual Comput. Graphics}, 28(1):1073--1083, 2022. \href{https://doi.org/10.1109/TVCG.2021.3114770}
{doi: {{%
10\hspace{.1pt}\discretionary{.}{%
}{.}\hspace{.4pt}1109\discretionary{/}{%
}{/}TVCG\hspace{.1pt}\discretionary{.}{%
}{.}\hspace{.4pt}2021\hspace{.1pt}\discretionary{.}{%
}{.}\hspace{.4pt}3114770}}}


\bibitem{Mackinlay1986-AutoDesign}
J.~Mackinlay.
\newblock Automating the design of graphical presentations of relational information.
\newblock {\em ACM Trans. Graph.}, 5(2):110–141,  32 pages, apr 1986. \href{https://doi.org/10.1145/22949.22950}
{doi: {{%
10\hspace{.1pt}\discretionary{.}{%
}{.}\hspace{.4pt}1145\discretionary{/}{%
}{/}22949\hspace{.1pt}\discretionary{.}{%
}{.}\hspace{.4pt}22950}}}


\bibitem{markant2023-viz_persuade}
D.~Markant, M.~Rogha, A.~Karduni, R.~Wesslen, and W.~Dou.
\newblock When do data visualizations persuade? the impact of prior attitudes on learning about correlations from scatterplot visualizations.
\newblock In {\em Proc. CHI},  article no. 174,  16 pages. ACM, New York, 2023. \href{https://doi.org/10.1145/3544548.3581330}
{doi: {{%
10\hspace{.1pt}\discretionary{.}{%
}{.}\hspace{.4pt}1145\discretionary{/}{%
}{/}3544548\hspace{.1pt}\discretionary{.}{%
}{.}\hspace{.4pt}3581330}}}


\bibitem{markant2022-attitude_change_minds}
D.~B. Markant, M.~Rogha, A.~Karduni, R.~Wesslen, and W.~Dou.
\newblock Can data visualizations change minds? identifying mechanisms of elaborative thinking and persuasion.
\newblock In {\em 2022 IEEE Workshop on Visualization for Social Good (VIS4Good)}, pp. 1--5, 2022. \href{https://doi.org/10.1109/VIS4Good57762.2022.00005}
{doi: {{%
10\hspace{.1pt}\discretionary{.}{%
}{.}\hspace{.4pt}1109\discretionary{/}{%
}{/}VIS4Good57762\hspace{.1pt}\discretionary{.}{%
}{.}\hspace{.4pt}2022\hspace{.1pt}\discretionary{.}{%
}{.}\hspace{.4pt}00005}}}


\bibitem{McNutt2020-Mirages}
A.~McNutt, G.~Kindlmann, and M.~Correll.
\newblock Surfacing visualization mirages.
\newblock In {\em In Proc. CHI},  16 pages, p. 1–16. ACM, New York, 2020. \href{https://doi.org/10.1145/3313831.3376420}
{doi: {{%
10\hspace{.1pt}\discretionary{.}{%
}{.}\hspace{.4pt}1145\discretionary{/}{%
}{/}3313831\hspace{.1pt}\discretionary{.}{%
}{.}\hspace{.4pt}3376420}}}


\bibitem{miles1994-qualitativeAnalysis}
M.~B. Miles and A.~M. Huberman.
\newblock {\em Qualitative data analysis: An expanded sourcebook}.
\newblock sage, 1994.

\bibitem{nanayakkara2022-dpTradeoffs}
P.~Nanayakkara, J.~Bater, X.~He, J.~Hullman, and J.~Rogers.
\newblock Visualizing privacy-utility trade-offs in differentially private data releases.
\newblock {\em Proceedings on Privacy Enhancing Technologies}, 2022(2), 2022. \href{https://doi.org/10.2478/popets-2022-0058}
{doi: {{%
10\hspace{.1pt}\discretionary{.}{%
}{.}\hspace{.4pt}2478\discretionary{/}{%
}{/}popets\discretionary{%
}{-}{-}2022\discretionary{%
}{-}{-}0058}}}


\bibitem{nurgalieva2019-info_need_older_adults}
L.~Nurgalieva, A.~Frik, F.~Ceschel, S.~Egelman, and M.~Marchese.
\newblock Information design in an aged care context: Views of older adults on information sharing in a care triad.
\newblock In {\em Proc. PervasiveHealth}, PervasiveHealth'19,  10 pages, p. 101–110. ACM, New York, 2019. \href{https://doi.org/10.1145/3329189.3329211}
{doi: {{%
10\hspace{.1pt}\discretionary{.}{%
}{.}\hspace{.4pt}1145\discretionary{/}{%
}{/}3329189\hspace{.1pt}\discretionary{.}{%
}{.}\hspace{.4pt}3329211}}}


\bibitem{oscar2017-personal_viz}
N.~Oscar, S.~Mej\'{\i}a, R.~Metoyer, and K.~Hooker.
\newblock Towards personalized visualization: Information granularity, situation, and personality.
\newblock In {\em Proc. DIS}, p. 811–819. ACM, New York, 2017. \href{https://doi.org/10.1145/3064663.3064704}
{doi: {{%
10\hspace{.1pt}\discretionary{.}{%
}{.}\hspace{.4pt}1145\discretionary{/}{%
}{/}3064663\hspace{.1pt}\discretionary{.}{%
}{.}\hspace{.4pt}3064704}}}


\bibitem{Padilla2018-framework}
L.~Padilla, S.~Creem-Regehr, M.~Hegarty, and J.~K. Stefanucci.
\newblock Decision making with visualizations: A cognitive framework across disciplines.
\newblock {\em Cogn. Research}, 3(29), 2018. \href{https://doi.org/10.1186/s41235-018-0120-9}
{doi: {{%
10\hspace{.1pt}\discretionary{.}{%
}{.}\hspace{.4pt}1186\discretionary{/}{%
}{/}s41235\discretionary{%
}{-}{-}018\discretionary{%
}{-}{-}0120\discretionary{%
}{-}{-}9}}}


\bibitem{padilla2016-binscalar}
L.~Padilla, P.~S. Quinan, M.~Meyer, and S.~H. Creem-Regehr.
\newblock Evaluating the impact of binning 2d scalar fields.
\newblock {\em IEEE Trans. Visual Comput. Graphics}, 23(1):431--440, 2017. \href{https://doi.org/10.1109/TVCG.2016.2599106}
{doi: {{%
10\hspace{.1pt}\discretionary{.}{%
}{.}\hspace{.4pt}1109\discretionary{/}{%
}{/}TVCG\hspace{.1pt}\discretionary{.}{%
}{.}\hspace{.4pt}2016\hspace{.1pt}\discretionary{.}{%
}{.}\hspace{.4pt}2599106}}}


\bibitem{panavas2023investigatingDiffPrivate}
L.~Panavas, T.~Crnovrsanin, J.~L. Adams, J.~Ullman, A.~Sargavad, M.~Tory, and C.~Dunne.
\newblock Investigating the visual utility of differentially private scatterplots.
\newblock {\em IEEE Trans. Visual Comput. Graphics}, 30(8):5370--5385, 2024. \href{https://doi.org/10.1109/TVCG.2023.3292391}
{doi: {{%
10\hspace{.1pt}\discretionary{.}{%
}{.}\hspace{.4pt}1109\discretionary{/}{%
}{/}TVCG\hspace{.1pt}\discretionary{.}{%
}{.}\hspace{.4pt}2023\hspace{.1pt}\discretionary{.}{%
}{.}\hspace{.4pt}3292391}}}


\bibitem{pandey2014-persuasive}
A.~V. Pandey, A.~Manivannan, O.~Nov, M.~Satterthwaite, and E.~Bertini.
\newblock The persuasive power of data visualization.
\newblock {\em IEEE Trans. Visual Comput. Graphics}, 20(12):2211--2220, 2014. \href{https://doi.org/10.1109/TVCG.2014.2346419}
{doi: {{%
10\hspace{.1pt}\discretionary{.}{%
}{.}\hspace{.4pt}1109\discretionary{/}{%
}{/}TVCG\hspace{.1pt}\discretionary{.}{%
}{.}\hspace{.4pt}2014\hspace{.1pt}\discretionary{.}{%
}{.}\hspace{.4pt}2346419}}}


\bibitem{price2016effects}
M.~M. Price, J.~J. Crumley-Branyon, W.~R. Leidheiser, and R.~Pak.
\newblock Effects of information visualization on older adults’ decision-making performance in a medicare plan selection task: a comparative usability study.
\newblock {\em JMIR human factors}, 3(1):e5106, 2016. \href{https://doi.org/10.2196/humanfactors.5106}
{doi: {{%
10\hspace{.1pt}\discretionary{.}{%
}{.}\hspace{.4pt}2196\discretionary{/}{%
}{/}humanfactors\hspace{.1pt}\discretionary{.}{%
}{.}\hspace{.4pt}5106}}}


\bibitem{pu2020-PGoG}
X.~Pu and M.~Kay.
\newblock A probabilistic grammar of graphics.
\newblock In {\em Proc. CHI}, CHI '20,  13 pages, p. 1–13. ACM, New York, 2020. \href{https://doi.org/10.1145/3313831.3376466}
{doi: {{%
10\hspace{.1pt}\discretionary{.}{%
}{.}\hspace{.4pt}1145\discretionary{/}{%
}{/}3313831\hspace{.1pt}\discretionary{.}{%
}{.}\hspace{.4pt}3376466}}}


\bibitem{sarma2022evaluatingMissingValue}
A.~Sarma, S.~Guo, J.~Hoffswell, R.~Rossi, F.~Du, E.~Koh, and M.~Kay.
\newblock Evaluating the use of uncertainty visualisations for imputations of data missing at random in scatterplots.
\newblock {\em IEEE Trans. Visual Comput. Graphics}, 29(1):602--612, 2023. \href{https://doi.org/10.1109/TVCG.2022.3209348}
{doi: {{%
10\hspace{.1pt}\discretionary{.}{%
}{.}\hspace{.4pt}1109\discretionary{/}{%
}{/}TVCG\hspace{.1pt}\discretionary{.}{%
}{.}\hspace{.4pt}2022\hspace{.1pt}\discretionary{.}{%
}{.}\hspace{.4pt}3209348}}}


\bibitem{satyanarayan2016-vegalite}
A.~Satyanarayan, D.~Moritz, K.~Wongsuphasawat, and J.~Heer.
\newblock Vega-lite: A grammar of interactive graphics.
\newblock {\em IEEE Trans. Visual Comput. Graphics}, 23(1):341--350, 2017. \href{https://doi.org/10.1109/TVCG.2016.2599030}
{doi: {{%
10\hspace{.1pt}\discretionary{.}{%
}{.}\hspace{.4pt}1109\discretionary{/}{%
}{/}TVCG\hspace{.1pt}\discretionary{.}{%
}{.}\hspace{.4pt}2016\hspace{.1pt}\discretionary{.}{%
}{.}\hspace{.4pt}2599030}}}


\bibitem{schetinger2023doom}
V.~Schetinger, S.~Di~Bartolomeo, M.~El-Assady, A.~McNutt, M.~Miller, J.~P.~A. Passos, and J.~L. Adams.
\newblock Doom or deliciousness: Challenges and opportunities for visualization in the age of generative models.
\newblock In {\em Comput. Graphics Forum}, vol.~42, pp. 423--435. Wiley Online Library, 2023. \href{https://doi.org/10.1111/cgf.14841}
{doi: {{%
10\hspace{.1pt}\discretionary{.}{%
}{.}\hspace{.4pt}1111\discretionary{/}{%
}{/}cgf\hspace{.1pt}\discretionary{.}{%
}{.}\hspace{.4pt}14841}}}


\bibitem{sharif2023-uncertain_screenreader}
A.~Sharif, R.~Zhong, and Y.~Wang.
\newblock Conveying uncertainty in data visualizations to screen-reader users through non-visual means.
\newblock In {\em Proc. ASSETS}, ASSETS '23. ACM, New York, 2023. \href{https://doi.org/10.1145/3597638.3614502}
{doi: {{%
10\hspace{.1pt}\discretionary{.}{%
}{.}\hspace{.4pt}1145\discretionary{/}{%
}{/}3597638\hspace{.1pt}\discretionary{.}{%
}{.}\hspace{.4pt}3614502}}}


\bibitem{shneiderman2003eyes}
B.~Shneiderman.
\newblock The eyes have it: a task by data type taxonomy for information visualizations.
\newblock In {\em Proceedings 1996 IEEE Symposium on Visual Languages}, pp. 336--343, 1996. \href{https://doi.org/10.1109/VL.1996.545307}
{doi: {{%
10\hspace{.1pt}\discretionary{.}{%
}{.}\hspace{.4pt}1109\discretionary{/}{%
}{/}VL\hspace{.1pt}\discretionary{.}{%
}{.}\hspace{.4pt}1996\hspace{.1pt}\discretionary{.}{%
}{.}\hspace{.4pt}545307}}}


\bibitem{simpson1951interpretation}
E.~H. Simpson.
\newblock The interpretation of interaction in contingency tables.
\newblock {\em Journal of the Royal Statistical Society: Series B (Methodological)}, 13(2):238--241, 1951.

\bibitem{sondag2017stableTreemap}
M.~Sondag, B.~Speckmann, and K.~Verbeek.
\newblock Stable treemaps via local moves.
\newblock {\em IEEE Trans. Visual Comput. Graphics}, 24(1):729--738, 2018. \href{https://doi.org/10.1109/TVCG.2017.2745140}
{doi: {{%
10\hspace{.1pt}\discretionary{.}{%
}{.}\hspace{.4pt}1109\discretionary{/}{%
}{/}TVCG\hspace{.1pt}\discretionary{.}{%
}{.}\hspace{.4pt}2017\hspace{.1pt}\discretionary{.}{%
}{.}\hspace{.4pt}2745140}}}


\bibitem{speckmann2010necklace}
B.~Speckmann and K.~Verbeek.
\newblock Necklace maps.
\newblock {\em IEEE Trans. Visual Comput. Graphics}, 16(6):881--889, 2010. \href{https://doi.org/10.1109/TVCG.2010.180}
{doi: {{%
10\hspace{.1pt}\discretionary{.}{%
}{.}\hspace{.4pt}1109\discretionary{/}{%
}{/}TVCG\hspace{.1pt}\discretionary{.}{%
}{.}\hspace{.4pt}2010\hspace{.1pt}\discretionary{.}{%
}{.}\hspace{.4pt}180}}}


\bibitem{sweeney2002-kanon}
L.~Sweeny.
\newblock k-anonymity: A model for protecting privacy.
\newblock {\em Int. J. Uncertainty Fuzziness Knowledge Based Syst.}, 10(05):557--570, 2002. \href{https://doi.org/10.1142/S0218488502001648}
{doi: {{%
10\hspace{.1pt}\discretionary{.}{%
}{.}\hspace{.4pt}1142\discretionary{/}{%
}{/}S0218488502001648}}}


\bibitem{trumbo1981-theoryBivariate}
B.~E. Trumbo.
\newblock A theory for coloring bivariate statistical maps.
\newblock {\em Am. Stat.}, 35(4):220--226, 1981. \href{https://doi.org/10.1080/00031305.1981.10479360}
{doi: {{%
10\hspace{.1pt}\discretionary{.}{%
}{.}\hspace{.4pt}1080\discretionary{/}{%
}{/}00031305\hspace{.1pt}\discretionary{.}{%
}{.}\hspace{.4pt}1981\hspace{.1pt}\discretionary{.}{%
}{.}\hspace{.4pt}10479360}}}


\bibitem{Vickers2013-CategoryTheory}
P.~Vickers, J.~Faith, and N.~Rossiter.
\newblock Understanding visualization: A formal approach using category theory and semiotics.
\newblock {\em IEEE Trans. Visual Comput. Graphics}, 19(6):1048–1061,  14 pages, jun 2013. \href{https://doi.org/10.1109/TVCG.2012.294}
{doi: {{%
10\hspace{.1pt}\discretionary{.}{%
}{.}\hspace{.4pt}1109\discretionary{/}{%
}{/}TVCG\hspace{.1pt}\discretionary{.}{%
}{.}\hspace{.4pt}2012\hspace{.1pt}\discretionary{.}{%
}{.}\hspace{.4pt}294}}}


\bibitem{wickham2010-graphicalLineups}
H.~Wickham, D.~Cook, H.~Hofmann, and A.~Buja.
\newblock Graphical inference for infovis.
\newblock {\em IEEE Trans. Visual Comput. Graphics}, 16(6):973--979, 2010. \href{https://doi.org/10.1109/TVCG.2010.161}
{doi: {{%
10\hspace{.1pt}\discretionary{.}{%
}{.}\hspace{.4pt}1109\discretionary{/}{%
}{/}TVCG\hspace{.1pt}\discretionary{.}{%
}{.}\hspace{.4pt}2010\hspace{.1pt}\discretionary{.}{%
}{.}\hspace{.4pt}161}}}


\bibitem{wilkinson2012grammar}
L.~Wilkinson.
\newblock The grammar of graphics.
\newblock In {\em Handbook of computational statistics: Concepts and methods}, pp. 375--414. Springer, 2011. \href{https://doi.org/10.1007/0-387-28695-0}
{doi: {{%
10\hspace{.1pt}\discretionary{.}{%
}{.}\hspace{.4pt}1007\discretionary{/}{%
}{/}0\discretionary{%
}{-}{-}387\discretionary{%
}{-}{-}28695\discretionary{%
}{-}{-}0}}}


\bibitem{windhager2024complexity}
F.~Windhager, A.~Abdul-Rahman, M.-J. Bludau, N.~Hengesbach, H.~Lamqaddam, I.~Meirelles, B.~Speckmann, and M.~Correll.
\newblock Complexity as design material position paper.
\newblock In {\em 2024 IEEE Evaluation and Beyond - Methodological Approaches for Visualization (BELIV)}, pp. 71--80, 2024. \href{https://doi.org/10.1109/BELIV64461.2024.00013}
{doi: {{%
10\hspace{.1pt}\discretionary{.}{%
}{.}\hspace{.4pt}1109\discretionary{/}{%
}{/}BELIV64461\hspace{.1pt}\discretionary{.}{%
}{.}\hspace{.4pt}2024\hspace{.1pt}\discretionary{.}{%
}{.}\hspace{.4pt}00013}}}


\bibitem{witt2019graph}
J.~K. Witt.
\newblock Graph construction: An empirical investigation on setting the range of the y-axis.
\newblock {\em Meta-psychology}, 2019.

\bibitem{wood2018design}
J.~Wood, A.~Kachkaev, and J.~Dykes.
\newblock Design exposition with literate visualization.
\newblock {\em IEEE Trans. Visual Comput. Graphics}, 25(1):759--768, 2019. \href{https://doi.org/10.1109/TVCG.2018.2864836}
{doi: {{%
10\hspace{.1pt}\discretionary{.}{%
}{.}\hspace{.4pt}1109\discretionary{/}{%
}{/}TVCG\hspace{.1pt}\discretionary{.}{%
}{.}\hspace{.4pt}2018\hspace{.1pt}\discretionary{.}{%
}{.}\hspace{.4pt}2864836}}}


\bibitem{Wu2024-transforms}
E.~Wu and R.~Cheng.
\newblock Design-specific transformations in visualization.
\newblock {\em IEEE VIS BELIV Workshop}, 2024. \href{https://doi.org/10.48550/arXiv.2407.06404}
{doi: {{%
10\hspace{.1pt}\discretionary{.}{%
}{.}\hspace{.4pt}48550\discretionary{/}{%
}{/}arXiv\hspace{.1pt}\discretionary{.}{%
}{.}\hspace{.4pt}2407\hspace{.1pt}\discretionary{.}{%
}{.}\hspace{.4pt}06404}}}


\bibitem{wu2023-data_data}
K.~Wu, M.~H. Tran, E.~Petersen, V.~Koushik, and D.~A. Szafir.
\newblock Data, data, everywhere: Uncovering everyday data experiences for people with intellectual and developmental disabilities.
\newblock In {\em Proc. CHI},  article no. 804,  17 pages. ACM, New York, 2023. \href{https://doi.org/10.1145/3544548.3581204}
{doi: {{%
10\hspace{.1pt}\discretionary{.}{%
}{.}\hspace{.4pt}1145\discretionary{/}{%
}{/}3544548\hspace{.1pt}\discretionary{.}{%
}{.}\hspace{.4pt}3581204}}}


\bibitem{wu2024-rationalAgentBenchmark}
Y.~Wu, Z.~Guo, M.~Mamakos, J.~Hartline, and J.~Hullman.
\newblock The rational agent benchmark for data visualization.
\newblock {\em IEEE Trans. Visual Comput. Graphics}, 30(1):338--347, 2024. \href{https://doi.org/10.1109/TVCG.2023.3326513}
{doi: {{%
10\hspace{.1pt}\discretionary{.}{%
}{.}\hspace{.4pt}1109\discretionary{/}{%
}{/}TVCG\hspace{.1pt}\discretionary{.}{%
}{.}\hspace{.4pt}2023\hspace{.1pt}\discretionary{.}{%
}{.}\hspace{.4pt}3326513}}}


\bibitem{xiong2022-iconarray}
C.~Xiong, A.~Sarvghad, D.~G. Goldstein, J.~M. Hofman, and C.~Demiralp.
\newblock Investigating perceptual biases in icon arrays.
\newblock In {\em Proc. CHI},  article no. 137,  12 pages. ACM, New York, 2022. \href{https://doi.org/10.1145/3491102.3501874}
{doi: {{%
10\hspace{.1pt}\discretionary{.}{%
}{.}\hspace{.4pt}1145\discretionary{/}{%
}{/}3491102\hspace{.1pt}\discretionary{.}{%
}{.}\hspace{.4pt}3501874}}}


\bibitem{yang2023-SubjectiveProb}
F.~Yang, M.~Hedayati, and M.~Kay.
\newblock Subjective probability correction for uncertainty representations.
\newblock In {\em Proc. CHI},  article no. 836,  17 pages. ACM, New York, 2023. \href{https://doi.org/10.1145/3544548.3580998}
{doi: {{%
10\hspace{.1pt}\discretionary{.}{%
}{.}\hspace{.4pt}1145\discretionary{/}{%
}{/}3544548\hspace{.1pt}\discretionary{.}{%
}{.}\hspace{.4pt}3580998}}}


\bibitem{Zhang2021_DiffPrivacy}
D.~Zhang, A.~Sarvghad, and G.~Miklau.
\newblock Investigating visual analysis of differentially private data.
\newblock {\em IEEE Trans. Visual Comput. Graphics}, 27(2):1786--1796, 2021. \href{https://doi.org/10.1109/TVCG.2020.3030369}
{doi: {{%
10\hspace{.1pt}\discretionary{.}{%
}{.}\hspace{.4pt}1109\discretionary{/}{%
}{/}TVCG\hspace{.1pt}\discretionary{.}{%
}{.}\hspace{.4pt}2020\hspace{.1pt}\discretionary{.}{%
}{.}\hspace{.4pt}3030369}}}


\end{thebibliography}
